\documentclass[a4paper,12pt]{article}
\pdfoutput=1

\usepackage[a4paper,top=2cm,bottom=2cm,left=3cm,right=2cm]{geometry}
\usepackage{amssymb,graphicx}
\usepackage{epstopdf}
\usepackage{amsmath,amsfonts}
\usepackage{mathtools}
\usepackage[colorlinks=true]{hyperref}
\hypersetup{
     colorlinks   = true,
     citecolor    = blue
}
\usepackage[dvipsnames]{xcolor}
\usepackage{caption}
\usepackage{subcaption}
\usepackage{cite}

\usepackage[T1]{fontenc}
\usepackage[utf8]{inputenc}

\def\d{\partial}

\newcommand{\be}{\begin{equation}}
\newcommand{\ee}{\end{equation}}
\newcommand{\bea}{\begin{eqnarray}}
\newcommand{\eea}{\end{eqnarray}}
\newcommand{\bg}{\begin{gather}}
\newcommand{\eg}{\end{gather}}
\newcommand{\bseq}{\begin{subequations}}
\newcommand{\eseq}{\end{subequations}}

\def\half{\frac{1}{2}}

\DeclareMathOperator{\tr}{tr}

\newcommand{\pt}{\partial}

\newcommand{\I}{\mathcal I}

\numberwithin{equation}{section}

\begin{document}

\vskip -0.5in
\begin{minipage}{6.5in}
\begin{flushleft}
 {\small CCTP-2017-4 \\ ITCP-IPP 2017/15}
\end{flushleft}
\end{minipage}\\

\begin{center}
{\large \bf \center  BLACK HOLE ELASTICITY AND \\[0.35cm] 
GAPPED TRANSVERSE PHONONS IN HOLOGRAPHY}

\vspace{1.cm}

\textbf{Lasma Alberte$^{a,1}$, Martin Ammon$^{b,2}$, Matteo Baggioli$^{c, 3}$, Amadeo Jim\'enez$^{b,4}$, \\[3mm]Oriol Pujol{\`a}s$^{d,5}$
}\\

\vspace{1.cm}
${}^a\!\!$ {\em {Abdus Salam International Centre for Theoretical Physics (ICTP)\\Strada Costiera 11, 34151, Trieste, Italy}}

\vspace{.2cm}
${}^b\!\!$ {\em Theoretisch-Physikalisches Institut, Friedrich-Schiller-Universit\"at Jena,
Max-Wien-Platz 1, D-07743 Jena, Germany
}

\vspace{.2cm}
${}^c\!\!$ {\em Crete Center for Theoretical Physics, Institute for Theoretical and Computational Physics, Department of Physics, University of Crete, 71003
Heraklion, Greece
}

\vspace{.2cm}
${}^d\!\!$ {\em 
{Institut de F\'isica d'Altes Energies (IFAE)\\ 
The Barcelona Institute of Science and Technology (BIST)\\
Campus UAB, 08193 Bellaterra (Barcelona) Spain
}
}

\end{center}

\vspace{0.8cm}

\centerline{\bf Abstract}
\vspace{2 mm}
\begin{quote}\small   

We study the elastic response of planar black hole (BH) solutions in a simple class of holographic models with broken translational invariance. We compute the transverse quasi-normal mode spectrum and the propagation speed of the lowest energy mode. We find that the speed of the lowest mode relates to the BH rigidity modulus as dictated by elasticity theory. This allows to identify these modes as transverse phonons---the pseudo Goldstone bosons of spontaneously broken translational invariance. 
In addition, we show that these modes have a mass gap controlled by an explicit source of the translational symmetry breaking.
These results provide a new confirmation that the BHs in these models do exhibit solid properties that become more manifest at low temperatures. Also, by the AdS/CFT correspondence, this allows to extend the standard results from the effective field theory for solids to quantum-critical materials.

 \end{quote}

\begin{flushleft}
\href{mailto:lalberte@ictp.it}{lalberte@ictp.it}\\
\href{mailto:martin.ammon@uni-jena.de}{martin.ammon@uni-jena.de}\\
\href{mailto:mbaggioli@physics.uoc.gr}{mbaggioli@physics.uoc.gr}\\
\href{mailto:amadeo.jimenez.alba@uni-jena.de}{amadeo.jimenez.alba@uni-jena.de}\\
\href{mailto:pujolas@ifae.es}{pujolas@ifae.es}
\end{flushleft}
 
\newpage

\tableofcontents

\section{Introduction}
Translational invariance is a fundamental symmetry of nature---of the elementary vacuum and dynamics---but it is {\em badly} broken in condensed matter. 
Real-world physical systems break this symmetry in various ways. Already a perfectly ordered lattice structure accommodates  phonon excitations, the collective vibrational excitations of the lattice. Their inevitable presence stems from the symmetry breaking pattern: in solid state,  the phonons can be understood as the  Goldstone bosons associated to translational symmetry breaking \cite{Leutwyler:1993gf,Leutwyler:1996er}. Impurities, defects and other forms of random disorder  provide an additional source of  translational symmetry breaking. Both effects  have a deep impact on the transport parameters of the material as well as on the phonon properties. An example of the interplay between various sources of translational symmetry breaking appears in (pinned) charge-density waves and in Wigner crystals,  see \emph{e.g.} \cite{Lubensky,Gruner:1994zz}.

The main goal of this work is to show that the  black holes in rather simple gravitational theories exhibit completely analogous behaviour, displaying transverse phonon\footnote{We recall that solid materials support both longitudinal as well as transverse (or shear) sound waves. The transverse sound waves are vibrational modes with material displacements that are orthogonal to the direction of propagation.} excitations whose properties are very sensitive to how the translational symmetry is broken. 
We shall focus on  background geometries with Anti-de-Sitter (AdS) asymptotics, in order to have a simple Conformal Field Theory (CFT) interpretation of the physics via the standard holographic AdS/CFT dictionary \cite{Maldacena:1997re}. The simplest class of holographic models that allow to include the effects of breaking the translational invariance and of momentum relaxation are of Holographic Massive Gravity type \cite{Vegh:2013sk,Blake:2013bqa,Andrade:2013gsa,Baggioli:2014roa}. The motivation to consider these models is that they are computationally more tractable since the background geometries are homogeneous. The analysis of the elastic properties of these models was initiated in \cite{Baggioli:2014roa,Alberte:2015isw,Alberte:2016xja}. Some of the outcomes of these works are that $i)$ these black hole solutions do exhibit a clear notion of elasticity, starting with a well-defined and computable modulus of rigidity; $ii)$ a certain limit was identified in  \cite{Baggioli:2014roa} where the lowest quasi-normal mode (QNM) could be identified as a {\em bona fide} transverse pseudo phonon. 

In this work we will show that the latter property is indeed realized. This confirms that the Holographic Massive Gravity models do indeed incorporate the elastic response. These models can therefore be interpreted as CFTs with elastic properties or, equivalently, as elastic materials in a critical scale-invariant regime. 
Along the way we will uncover some interesting properties of these modes, such as their ability to acquire a mass-gap and what determines it.

\subsection{Transverse phonons}
Let us now explain what we mean by {\em bona fide} phonons. From the low-frequency point of view, the hallmark of phonons is in their propagation sound speeds. Indeed, it is well known that in solids with negligible damping and mass-gap generating effects the phonon dispersion relations are linear,
\begin{equation}\label{disp}
\omega\,=\,c_{L,\,T}\;k\,,
\end{equation}
and the longitudinal and transverse sound speeds, $c_{L,\,T}$, are given by (in $3+1$ dimensions):
\begin{equation}\label{speeds}
c_L^2\,=\,\frac{\kappa\,+\,\frac{4}{3}\,\mu}{\varepsilon+p}\,,\qquad c_T^2\,=\,\frac{\mu}{\varepsilon+p}
\end{equation}
where $\kappa$ and $\mu$ are the bulk (or compression) and shear (or rigidity) moduli, and  $\varepsilon$, $p$ stand for the energy density and the pressure. Most textbooks \cite{landau7,Lubensky,Leutwyler:1993gf, Leutwyler:1996er} derive these relations in the non-relativistic limit where the denominator simplifies to just $\varepsilon$. One can see in various ways that the extension to the relativistic case proceeds by the replacement $\varepsilon \to \varepsilon+p$, as is familiar from relativistic hydrodynamics.\footnote{See also \cite{Delacretaz:2017zxd} for a recent discussion in the case of spontaneous and explicit breaking of translational invariance.}

We give another derivation of \eqref{speeds} from the Effective Field Theory (EFT) for Solids \cite{Leutwyler:1993gf,Leutwyler:1996er,Dubovsky:2011sj,Endlich:2012pz,Nicolis:2013lma} in  Appendix \ref{app:eft}. Let us emphasize here that an illuminating aspect of the EFT of Solids is that it makes it very clear that the connection between the phonon velocities and the elastic moduli stems from symmetry principles.

In the present work we focus on transverse phonons. Solids (\emph{i.e.} materials with a non-zero rigidity modulus $\mu$) stand out as the materials that support propagating shear or transverse waves. Therefore the black hole geometries with a natural notion of the rigidity modulus \cite{Alberte:2015isw} offer a possibility to realize transverse phonons in the Goldstone-boson sense that can be tested by the relation \eqref{speeds}. The main result of this work is to show in  what limit of the Holographic Massive Gravity (HMG) models the Eq. \eqref{speeds} gives a very good approximation for the transverse (pseudo) phonon QNMs speed, with the natural notion of elastic modulus introduced in \cite{Alberte:2015isw}. This justifies that these QNMs can be identified as transverse pseudo phonons. In turn, this confirms that the black brane backgrounds that we consider do have elastic properties akin to the standard solids. In the CFT interpretation of the same events, what we shall describe is a transverse phonon in a critical solid material.

\subsection{Gapping}

The only obstruction in realizing holographically  transverse phonons is that the physics of these modes is actually much richer than simply Eq. \eqref{speeds}. Being Goldstone bosons, they are also sensitive to any {\em explicit} source of the translational symmetry breaking. 
This is analogous to  the spontaneous breaking of the chiral symmetry by the quark $\langle\bar q\, q \rangle$ condensate in QCD  and the associated appearance of pions as the Goldstone bosons. In this case the quark masses give rise to an additional explicit breaking of the chiral symmetry and this gives the would-be massless pions a mass. 

For translational symmetry, what an "explicit breaking" means is a bit more subtle and perhaps confusing to some readers. The physically  relevant  thing is the presence of another source of translational symmetry breaking, the consequence of which is that the transverse phonons acquire a mass gap.
In the real world, systems with a clear notion of multiple breakings of translational symmetry exist, perhaps the sharpest examples being  the pinned charge-density waves and Wigner crystals. 
In the models presented here there will be a clear notion of distinct sources of translational symmetry breaking---in the CFT parlance, the breakdown will take place differently at different energy scales.
In the present work we shall step aside from the discussion of what is the corresponding condensed matter phenomenon that fits best as a CFT interpretation of how the transverse phonons become gapped in our models. We prefer to stick to a more agnostic discussion in terms of symmetries only. The net effect, in any case, is that the degrees of freedom that we will find approximately do obey \eqref{speeds} and are gapped in the simplest models analyzed here. By analogy with pions, it is fair enough to say that this is due to an additional {\em explicit} source of the breaking of translational invariance which appears to be inevitable in our setup.  

Going back to the matter: in what limit should we expect to find modes that can be identified as transverse phonons? The answer is obvious by again looking at the EFT describing  QCD at low energies. The (almost tautological) observation is that the properties of the Goldstone bosons are  protected provided that the explicit symmetry breaking source is small enough. Hence, a comparison between two scales is required (which is what makes the observation not tautological):
the explicit breaking scale vs. the dynamical scale in the problem. For QCD this is $m_q / \Lambda_{QCD}$. Provided $m_q / \Lambda_{QCD} \ll 1$, the pions still keep their  pseudo-Goldstone boson properties. This suggests that in order to possibly realize {\em bona fide} transverse phonons we need to insure that the explicit breaking source is {\em weak} (small), so that the phonon mass gap is small compared to the typical energy scale of the problem.\footnote{In the opposite limit with large explicit breaking, there is no light scalar mode, and the sound speed of the lightest mode does not need to obey \eqref{speeds}.} 

In the first examples of HMG models \cite{Vegh:2013sk,Andrade:2013gsa} 
the explicit breaking is large (comparable to the spontaneous) and this translates into the absence of clearly identifiable propagating phonon excitations in the shear channel. As we show below, the situation  is reversed when considering HMG models with a metric potential $V(X)$ of the form \cite{Baggioli:2014roa}
\be\label{bench}
V(X) = X + \beta \, X^N
\ee
(we refer to Section \ref{sec:solids} for the precise definitions of $V$ and $X$) with large enough values for the  parameters $\beta$ and $N$.\footnote{The model in Ref \cite{Andrade:2013gsa} corresponds to $\beta=0$, and those of \cite{Vegh:2013sk} to  $N=1/2$. These values of $N$ and $\beta$ are too small.} We shall exhibit that this is indeed the case for the representative benchmark model with $N=5$~\footnote{Let us emphasize here that we expect our results to hold for potentials $V(X)$ more  general than \eqref{bench}, so long as they are monotonous and their behaviour at small (large) $X$ coincides with that in \eqref{bench}. For this reason, we view  \eqref{bench} simply as a convenient way to classify when do light propagating transverse phonons appear rather than a truncation of a series in powers of $X$.
} .

As shown below, for $\beta \gg1$ this model exhibits a gapped (but {\em light}) transverse phonon QNM with the following properties: \\

$i)$ the mass gap $\omega_0$ (or, equivalently, the {\em pinning frequency}) is parametrically smaller than the typical scale $\Lambda$ of the system, $\omega_0 \ll \Lambda$. Let us emphasize that the 'typical' scale $\Lambda$ should be an {\em intrinsic} scale in the problem. This excludes the temperature $T$ from playing that role, as it should be since we expect that the solid behaviour melts down at too high $T$.
{Having again QCD in mind, a natural reference value for $\Lambda$ is the typical separation in the frequency plane of the QNMs (}at low temperature). As a proxy for this we shall take the absolute value of the second QNM, that is, 
$\Lambda \equiv |\omega_2|$.\footnote{We reserve $\omega_1$ for the lowest QNM---the light phonon pole. Let us also add that the scale of the QNM mass separation, $|\omega_2|$,  
 is typically comparable to the energy scale obtained from the entropy density.} 
Thus, a good measure of how close to the pure Goldstone limit we are is the ratio
\be\label{indicator}
\frac{\omega_0}{\Lambda}~.
\ee
In our benchmark models, this ratio is controlled by the parameter $\beta$ and can be made arbitrarily small. Consequently, we obtain a parametrically {\em light} transverse phonon QNM.\\

$ii)$ the propagation speed of this mode complies with \eqref{speeds}. The agreement between the two quantities on the left and right hand sides of \eqref{speeds} is, of course, not perfect because of the explicit breaking---the mass gap. Our phonons are pseudo-Goldstone bosons, and therefore their properties have to coincide with the pure gapless Goldstone limit \eqref{speeds} only within a certain precision. 
The same logic suggests that a good indicator of the precision with which relations like \eqref{speeds} should hold is precisely the measure of how close to purely spontaneous case we are. This reduces to positive powers of \eqref{indicator}.
Our numerical results are completely in line with this picture.\\

\begin{figure}
\center
\includegraphics[width=0.48\textwidth]{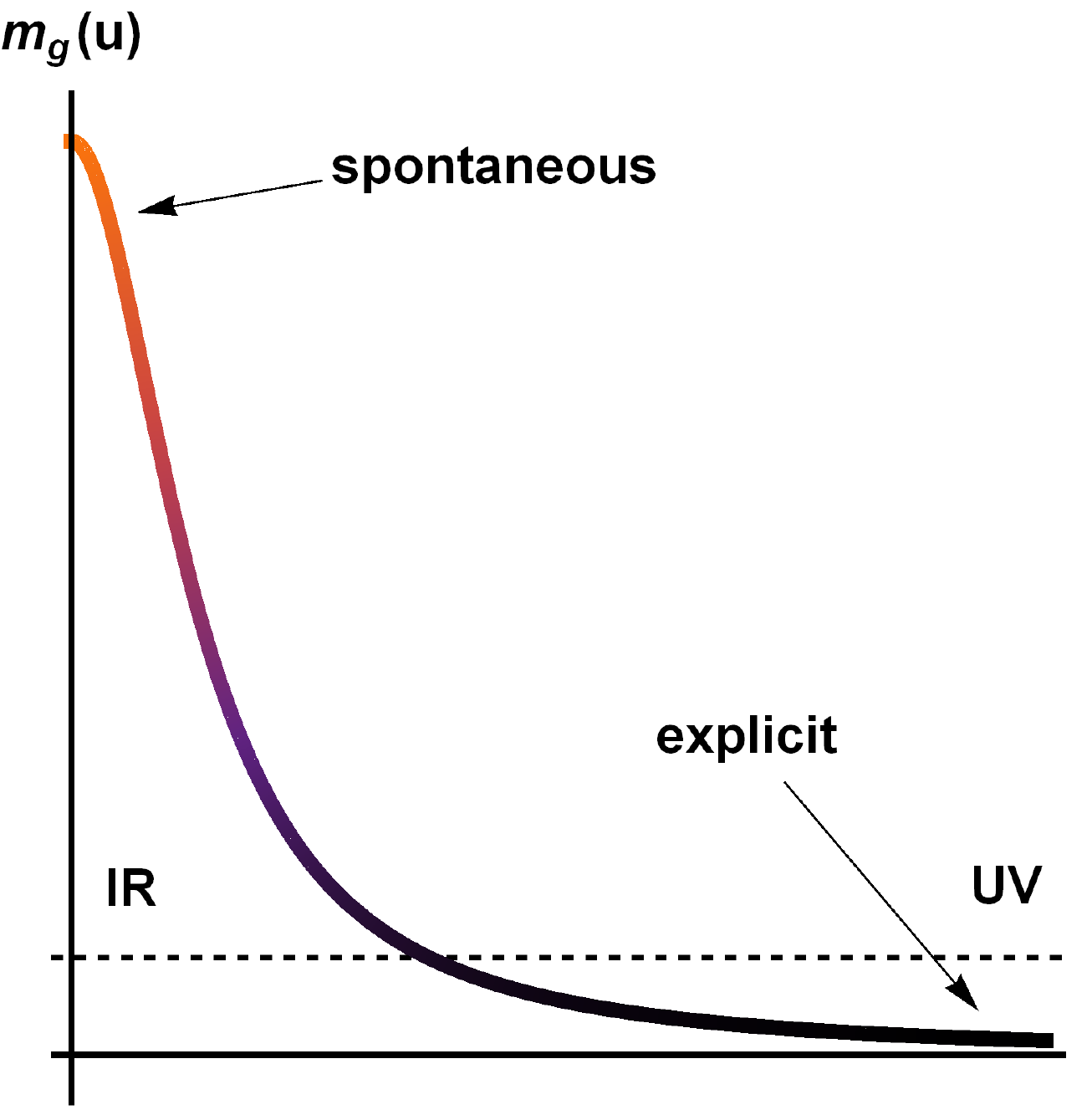}
\caption{A sketch of the behaviour of the graviton mass as a function of the holographic  coordinate $u$. The dashed line represents the mass profile  in the models  \cite{Andrade:2013gsa}. The colored line is the setup which we  analyze in this work and where we would expect a propagating gapped phonon to appear in the spectrum with a small mass gap $\omega_0$ controlled by the UV graviton mass. In practice, one can achieve this hierarchy using the HMG models of the form \eqref{bench} with large enough $N$ and $\beta$ \cite{Baggioli:2014roa}.}
\label{fig1}
\end{figure}

Let us sketch why  a HMG model like the one given in \eqref{bench} manages to generate light transverse phonons as seen from  the gravity dual picture. As explained in \cite{Baggioli:2014roa}, the only gauge invariant notion of an order parameter for the translational symmetry breaking in the presence of dynamical gravity is the graviton mass term, $m_g$. For generic potentials $V(X)$, one easily finds that  $m_g \propto V'(X)$. The important point is that on the background solution this graviton mass term depends on the holographic direction. Therefore, in general, there is an evolution of the amount of the symmetry breaking as we move along the radial direction. In the CFT language, this translates into the fact that along the RG flow generated by these deformations the amount of the breaking of translations can vary. This, however, only happens provided that $V'(X)$ is non-constant.

It is clear then that by an appropriate choice of the function $V(X)$ one can either break the translations more in the UV or in the IR---there are definitely distinct sources of symmetry breaking. The benchmark models \eqref{bench}, with large enough $N$ and $\beta$, are designed to break the translations most strongly in the IR region of the flow, with a separable and smaller breaking taking place in the UV region. This corresponds to  a graviton mass profile that peaks in the IR region, as depicted in Fig.~\ref{fig1}. By analogy with the standard terminology in AdS/CFT, we shall refer to the breaking associated to the UV region as {\em explicit}, and the one in the IR as {\em spontaneous}. See Sec. \ref{sec:SB} for  more details on this point.

\subsection{Damping}

An inescapable consequence of the smallness of the phonon mass gap is that they are easily excitable and therefore also very sensitive to thermal and other dissipative effects. Indeed, at temperatures as small as the mass gap thermal effects are expected to  be noticeable in the phonon dispersion relation, $\omega^2 = c_T^2 k^2 + \omega_0^2$, with both $\omega_0$ and $c_T$ possibly having  temperature corrections.
Since this can jeopardize the correct identification of the phonon sound speed from the dispersion relation, it is important to understand and isolate these dissipative effects.

Fortunately, there are two facts that will help in this task:

-- First, the analytic structure of the retarded Green's function relevant for the transport properties constrains quite a lot the way the dissipative effects enter in the final dispersion relation for the physical phonons (that are identified as QNMs). To a very good approximation, we find that the poles in the complex frequency plane $\omega$ for the lightest QNMs follow simply from the standard phonon wave equation corrected by a mass term and a friction term:
$$
\left[\partial_t^2 + \Gamma \,\partial_t- c_T^2 \, \partial_i^2 - \omega_0^2\right] \phi^{(T)} =0.
$$
The parameters ($\Gamma$, $c_T$, $\omega_0$) are real-valued and depend on temperature and other control parameters ($\beta$, etc.).  In Fourier space this leads to the simple quadratic equation
\begin{equation}\label{disp2}
\omega^2\,+\,i\,\Gamma\,\omega\,=\,\omega_0^2\,+\,c_T^2\,k^2\,.
\end{equation}
It is clear that $\Gamma$ is the only parameter encoding the dissipative effects and that it can be understood as an absorption or attenuation rate.

We also note that whenever both  $\omega_0$ and $\Gamma$ are small compared to the scale of the system, $\Lambda$, the $\Gamma $ and $\omega_0$ terms in the wave equation are small perturbations of the phonon degree of freedom introduced above. 
Indeed, for $\frac{\omega_0}{\Lambda},\frac{\Gamma}{\Lambda}\ll1$ there is a regime in momentum, $ \omega_0,\Gamma \ll c_T k \ll \Lambda$, for which the mode is propagating with an almost constant sound speed $d\omega/dk = c_T$. 
In this situation the inhomogeneous elastic transverse response is dominated by a sharp transverse phonon collective mode.\footnote{The presence of a sharp collective mode approaching the real axes is also evident in the optical conductivity properties of the model once a finite charge density is added \cite{Baggioli:2014roa}.}

At $k=0$, however, the location of the phonon poles is very sensitive to the comparison between $\Gamma$ and $2\omega_0$.
Indeed, the two roots of \eqref{disp2} read
\begin{equation}
\omega\,=\,\pm\,\sqrt{\omega_0^2\,-\,\frac{1}{4}\,\Gamma^2}\,-\,\frac{i}{2}\,\Gamma
\end{equation}
and exhibit a characteristic crossover between the under-damped and over-damped regimes (the latter corresponding to the eigenfrequencies having no real part). The crossover takes place whenever $\Gamma = 2\omega_0$ and is visualized as a characteristic `collision' \cite{Baggioli:2014roa,Davison:2014lua} of two purely imaginary poles which after the collision go into a pair of (propagating) poles with a non-zero  real part.\footnote{Note that when the real part of the QNM frequency does not vanish the absolute value $|\omega|$ coincides with $\omega_0$, which is what we identify as the gap. }

-- Second, the dissipative effects  (in our notation, $\Gamma$) must reduce at low temperatures. 
Since $\Gamma$ controls whether this mode is over/under damped, a good enough criterion to ensure that the dissipative effects are small is to require that $\Gamma \ll \omega_0$. 
The black hole solutions that we shall consider here do have a well-defined extremal zero-temperature regime in which the phonon modes are still guaranteed to exist by the symmetry breaking pattern. We find by direct numerical computations that in our black hole solutions we can lower enough $T$, so that $\Gamma \ll \omega_0$ is satisfied for the lowest QNM. The sound speed $c_T$ is then easily extracted by looking, {\em e.g.}, at the real part of its frequency. From this exercise we can test whether the relation \eqref{speeds} holds, and a good agreement is found.

Note that this behaviour of the quasi-normal shear modes is conceptually very  different from the one seen in hydrodynamics (or just the thermodynamics) of viscous fluids with momentum relaxation.\footnote{Nevertheless, by using a gradient expansion in small frequency $\omega$ and momentum $k$, one can infer the form of the hydrodynamic mode even in the presence of an explicit breaking. See \cite{Delacretaz:2017zxd} for a recent work on the topic.} 
This stark contrast is of course completely expected, because the black hole solutions that we study here have all the properties of solid materials, the dynamics of which differ from that of fluids. This difference is seen only at low temperatures, as expected, since typically solids do melt down into a fluid phase at a high enough temperature.  

For the black hole solutions studied here, the interplay between the solid behaviour (in the phonon dispersion relation) and the fluid behaviour is actually continuous:  the parameters $\Gamma$ and $\omega_0$ vary continuously with the external control parameters, going through and out of the over-damped regime. Therefore these BH solutions behave like solids that 'melt' at high temperatures in a continuous transition.\footnote{This is in line with our previous studies \cite{Alberte:2016xja} that suggest that these black hole solutions can be interpreted as viscoelastic materials.} We shall leave for the future a thorough investigation of this transition.  Here we limit ourselves to presenting the regime in which the solid behaviour, with propagating transverse phonons complying with \eqref{speeds}, is exhibited.

Let us emphasize that our interest lies in a completely opposite region with respect to the so-called  \textit{incoherent limit}\cite{Kim:2014bza,Davison:2014lua,Baggioli:2017ojd,Hartnoll:2014lpa,Grozdanov:2015qia,Davison:2015taa,Grozdanov:2015djs,Blake:2016sud,Baggioli:2016pia,Kim:2017dgz,Blake:2017qgd,Lucas:2015vna,Davison:2015bea,Lucas:2015lna,Baggioli:2016oqk},  where the momentum relaxation rate is parametrically smaller than the typical energy scale $\tau_{rel}^{-1}\ll T$. Indeed, in that regime momentum dissipates rapidly and no propagating degree of freedom survives; physics is just dominated by diffusion. On the contrary, for our setup having a small relaxation rate is crucial in order to qualify the pseudo-phonon mode as a proper propagating excitation.\\

\subsection*{Context}

There have been a large number of results in the context of holographic models that include momentum relaxation \cite{Vegh:2013sk,Davison:2013jba,Blake:2013bqa,Blake:2013owa,Davison:2013txa,Donos:2013eha,Andrade:2013gsa,Gouteraux:2014hca,Donos:2014uba,Amoretti:2014zha,Donos:2014oha,Baggioli:2014roa,Horowitz:2012ky,Baggioli:2016rdj,Baggioli:2016oju,Amoretti:2016cad} in the direction of mimicking and understanding the physics of the explicit breaking of translational invariance. As a counterpart, the spontaneous breaking of translations and, in particular, the study of the phonon modes in holography is still lacking a concrete playground and robust results. We hope that this work contributes to clarify some aspects of this debate.

Within the gauge--gravity duality framework the study of the Goldstone modes has been so far mainly focused on the spontaneous breaking of the $U(1)$ symmetry in relation to the onset of superfluidity and on the effects of Lorentz violation on the Goldstone theorem \cite{Amado:2013xya, Argurio:2015via,Esposito:2016ria}. The first study of the physcics of the pseudo-Goldstones in the context of strongly coupled theories has been initiated in \cite{Argurio:2015wgr} and \cite{Argurio:2016xih}. Just recently, the Ward identities for phonons and pseudo-phonons have been analyzed in \cite{Amoretti:2016bxs}, but without an explicit and computable model. One possible way of breaking spontaneously the translational invariance in the context of holography relies on the existence of a possible finite momentum instability of the Reissner-Nordstrom AdS black hole \cite{Nakamura:2009tf,Ooguri:2010kt,Aperis:2010cd,Ooguri:2010xs,Donos:2011bh,Bergman:2011rf,Donos:2011qt,Donos:2012wi,Ammon:2016szz}. In those situations, new ground states characterized by spontaneous lattice structures appear and the dual fied theory is usually related to the physics of striped materials and charge density waves. Despite all the existing models available, a precise identification of the collective phonon mode has not been accomplished yet.

In addition to our fundamental question of obtaining phonon's physics and elastic theory within the holographic framework, we can mention another two orthogonal and very interesting motivations in the context of holography and strongly coupled materials.

$i)$ Recently in \cite{Delacretaz:2016ivq}, it has been proposed that the transport properties, in particular the electric conductivity, of the bad metals exhibiting linear in $T$ resistivity could be explained by the interplay of explicit and spontaneous breaking of translational symmetry. The analysis has been performed within the framework of hydrodynamics and lacks a concrete model able to check the expectations and provide microscopic information on the various transport coefficients appearing in the hydrodynamic expansion. Our model is certainly a promising candidate for such a test.

$ii)$ Effective field theories for the spontaneous breaking of Poincar\'e symmetry have been developed in the context of weak coupling and flat space in \cite{Dubovsky:2011sj,Nicolis:2015sra,Nicolis:2013lma} using a language very similar to ours. It is a very valuable question and direction to understand how to embed them into the gauge--gravity duality. The discussions initiated in \cite{Alberte:2015isw} and our computations provide a first step towards that task.

\vspace{0.5cm}
The manuscript is organized as follows. In Section~\ref{sec:waves} we introduce the theory of elasticity from the field theory and the holographic sides and we show how to extract the shear modulus in holography. In Section~\ref{sec:solids} we present our holographic model for solids. In Section~\ref{sec:qnms} we analyze the spectrum of the transverse excitations of the system, we identify the gapped and damped transverse phonon, and we study in detail the speed of shear sound matching the results obtained from the shear modulus. Finally, in Section~\ref{sec:discussion} we discuss our results and conclude. In Appendix~\ref{app:eft} we review the EFT for phonons from a modern perspective and in \ref{app:numerics} we provide more details on the numerical analysis and our computations.

\section*{\sc Note added}
While this work was being finished, Refs. \cite{Jokela:2017ltu} and  \cite{Andrade:2017cnc} discussing similar issues in different holographic models were also being completed.

\section{Elasticity}\label{sec:waves}
In the standard mechanical linear response theory \cite{landau7,Lubensky} the deformation of a solid due to applied stress is described in terms of the displacement vector $u_i=x'_i-x_i$. The coordinates $x_i$ and $x'_i$ correspond to the coordinates of material points in the body before and after the deformation respectively. Under applied forces the distance $dl$ between different points changes to $dl'$ parametrized as
\be
dl'^2\equiv\delta_{ij}dx'_idx'_j=(\delta_{ij}+2u_{ij})dx_idx_j\,,
\ee
where $u_{ij}$ is the strain tensor defined as
\be\label{strain_tensor}
u_{ij}=\frac{1}{2}\left(\pt_i u_j+\pt_j u_i\right)\;.
\ee
This deformation brings the material out of its original state of equilibrium and creates internal stress trying to bring the body back to it. Due to the short range of the molecular forces creating the internal stress, the total force acting on any part of the solid body can be written as a surface integral: $\oint dS_j\sigma _{ij}$, where $\sigma_{ij}$ is the stress tensor.

\subsection{Elastic sound waves}
The quantity of interest for this work is the speed of propagation of elastic deformations in solids. The sound speed in an elastic medium is determined by its elastic properties. For homogeneous and isotropic solids (that is, upon coarse-graining on the microscopic structure) the linear elastic response parameters are the {\em shear elastic modulus}, $\mu$, and the {\em compression modulus}, $\kappa$. These relate the stress tensor $\sigma_{ik}$ to the strain tensor $u_{ik}$ as
\be\label{sigma1}
\sigma_{ik}\equiv\frac{\partial \mathcal F}{\partial u_{ik}}=\kappa u_{j}^j\delta_{ik}+2\mu(u_{ik}-\frac{1}{d}u_{j}^j\delta_{ik})\,,
\ee
where $\mathcal F$ is the free energy of the medium. The bulk modulus $\kappa$ describes the change in the volume of the body while the shear modulus $\mu$ describes the change in the shape of the body. For a pure shear deformation, $\kappa=0$, and only the traceless part of the stress tensor is non-zero, so that 
\begin{equation}\label{sigma2}
\sigma_{ik}^{(\text{T})}=2\mu \,u_{ik}^{(\text{T})}\,.
\end{equation} 
The Newtonian equations of motion for the propagation of small deformations inside the elastic medium are simply given by
\be
\rho \,\ddot u_i=\frac{\partial\sigma_{ik}}{\partial x_k}\,,
\ee
where $\rho$ is the mass density of the body. For a shear deformation in one particular direction the above equations reduce to plane wave equations for elastic waves propagating either in the direction of deformation (longitudinal) or in a direction perpendicular to the deformation (transverse).

The speeds of propagation of the two types of waves in relativistic systems are known to be: 
\be\label{sound_rel}
c^2_{{T}}=\frac{\mu}{\varepsilon+p}\,,\qquad c_L^2=\frac{\kappa+\frac{4}{3}\mu}{\varepsilon+p}\,.
\ee
For fluids, the modulus of rigidity $\mu$ is equal to zero so that  $c_T^2=0$ and only longitudinal sound waves propagate. The derivation of these expressions can be obtained via standard hydrodynamic methods \cite{KADANOFF}. We present another derivation by the standard EFT methodology for describing solid and fluid materials \cite{Leutwyler:1993gf, Leutwyler:1996er, Dubovsky:2011sj,Nicolis:2015sra,Nicolis:2013lma} in Appendix~\ref{app:eft}. We will make an explicit use of the expression \eqref{sound_rel} in the following.

\subsection{Elastic response in holography}
Another way to think about the displacement $u_i$ is to consider it as an infinitesimal diffeomorphism under which the coordinates transform to $x_i\to x'_i=x_i+u_i$. This leads to a transformation of spatial metric perturbations as
\be
h_{ij}\to h'_{ij}=h_{ij}-\partial_iu_j-\partial_ju_i\,.
\ee
Assuming that the coordinate system $x'_i$ was such that the metric perturbations were vanishing, \emph{i.e.} $h'_{ij}=0$, allows to make the identification
$h_{ij}=2u_{ij}$ 
and thus associate the strain tensor to metric perturbations. This identification is particularly useful in the context of holography. According to the standard prescription, the stress-energy tensor in the dual theory can be obtained by the variation of the renormalized Euclidean boundary action 
\cite{deHaro:2000vlm,Hartnoll:2009sz}: 
\be\label{emt1}
\left\langle T_{ij}\right\rangle=2\,\frac{\delta S_{\text{ren}}}{\delta\gamma^{ij}}=\frac{3}{2L}\,\gamma^{(3)}_{ij}\,,
\ee
where we have restricted to a $3+1$-dimensional bulk and set $8\pi G=1$. We define the near-boundary metric $\gamma_{\mu\nu}$ as
\be
\lim _{u\to 0} g_{\mu\nu}=\frac{L^2}{u^2}\gamma_{\mu\nu}\,,\qquad \gamma_{\mu\nu}=\gamma^{(0)}_{\mu\nu}+\left(\frac{u}{L}\right)^3\gamma^{(3)}_{\mu\nu}+\dots
\ee
The equation \eqref{emt1} can be written in the form of a linear response as
\be\label{emt2}
\left\langle T_{ij}\right\rangle={\cal G}^R_{T_{ij}\,T_{ij}} \gamma^{(0)}_{ij}
\ee
with the retarded Green's function given by
\be\label{gf}
{\cal G}^R_{T_{ij}\,T_{ij}} =\frac{3}{2L}\frac{\gamma^{(3)}_{ij}}{\gamma^{(0)}_{ij}}\,.
\ee
On the other hand, the expectation value of the stress tensor due to a deformation $u_{ik}$ on the boundary is given by 
\be\label{sigma3}
\left\langle\sigma_{ij}\right\rangle=\frac{\delta \mathcal F}{\delta u_{ij}}=-2\frac{\delta S_{\text{ren}}}{\delta \gamma_{ij}}=-\left\langle T_{ij}\right\rangle=-2{\cal G}^R_{T_{ij}\,T_{ij}} u_{ij}\,.
\ee
Restricting to the transverse traceless components and comparing to \eqref{sigma2} we find that the shear modulus can be expressed through the real part of the retarded Green's function as:
\be\label{mu}
\mu=-\text{Re }{\cal G}^R_{T_{ij}\,T_{ij}}\,.
\ee
From \eqref{sound_rel} we thus find that the transverse sound speed in relativistic dual solids can be expressed in terms of the coefficients in the near boundary expansion of metric perturbations as
\be\label{ct1}
c_T^2=-\frac{3}{2L}\frac{1}{\varepsilon+p}\text{Re }\frac{\gamma^{(3)}_{ij}}{\gamma^{(0)}_{ij}}\,.
\ee
In the following sections we shall determine the sound speed of the lowest quasi-normal modes in the transverse sector of the metric perturbations and compare to the expression above.

 We note that the shear modulus is contained in the real part of the Green's function, in contrast to the shear viscosity which lies in its imaginary part. The shear viscosity has been computed in holographic massive gravity models in \cite{Hartnoll:2016tri,Alberte:2016xja,Burikham:2016roo} and shows a violation of the Kovtun--Son--Starinets bound \cite{Policastro:2001yc}. In order to understand the transport properties in the shear sector dealing with viscoelasticity and in particular their definitions in terms of Kubo formulas in full detail one would need to perform a proper and generic hydrodynamic description of these models in the spirit of \cite{Blake:2015epa,Burikham:2016roo,Delacretaz:2017zxd}.

\section{Holographic Solids from massive gravity}\label{sec:solids}
We consider a $3+1$ dimensional gravity theory describing holographic solids, as introduced in \cite{Alberte:2015isw,Alberte:2016xja} (see also \cite{Baggioli:2014roa}):
\be\label{action}
S=\int d^4x\,\sqrt{-g}\left[\frac{1}{2}\left(R+\frac{6}{l^2}\right)-m^2\,V(X)\right]+\int_{u\to0}d^3x\sqrt{-\gamma}\,K\;,
\ee
where $l$ is the AdS radius, $m$ is a mass parameter of dimension $[m]=L^{-1}$, and
\begin{align}\label{XZI}
X  \equiv \half \tr[\I^{IJ}]\,,\qquad \mathcal I^{IJ}\equiv\d_\mu \phi^I \d^\mu \phi^J\,, 
\end{align}
and the indices $I,J=\{x,y\}$ are contracted with $\delta_{IJ}$. The last term in \eqref{action} is the Gibbons--Hawking boundary term where $\gamma$ is the induced metric on the AdS boundary, and $K=\gamma^{\mu\nu}\nabla_\mu n_\nu$ is the extrinsic curvature with $n^\mu$---an outward pointing unit normal vector to the boundary. On the scalar fields background $\hat\phi^I=\delta^I_ix^i$ the metric admits the black brane solution 
\be
\label{solmetric}
ds^2 =   \frac{l^2}{u^2}\left(\frac{du^2}{f(u)}-f(u)dt^2+dx^2+dy^2\right) \;,
\ee
so that the background value of $X=u^2/l^2$. Note that $X$ is dimensionless. The emblackening factor is given in terms of the background value of the mass potential as: 
\be\label{bf}
f(u)=1+u^3\int du\,\frac{m^2l^2}{u^4}V\left(\frac{u^2}{l^2}\right)-\tilde Mu^3\,,
\ee
where $\tilde M$ is a dimensionful integration constant given by the condition that $f(u_h)=0$. Henceforth we shall work in terms of the dimensionless length $z=u/u_h$. 
 The potential \eqref{bench} with $N=5$ on the background can be expressed as
\be\label{potential}
V(X)=X+\beta X^5=\left(\frac{u_h}{l}\right)^2z^2\left[1+\beta\left(\frac{u_h}{l}\right)^{8} z^{8}\right]\,.
\ee
The black brane temperature for the emblackening factor \eqref{bf} with the potential \eqref{potential} is then
\be\label{temp}
Tl = \frac{l}{u_h}\frac{|\partial_z f(1)|}{4\pi}=\frac{1}{4\pi }\left(\frac{l}{u_h}\right)\left[3-m^2u_h^2-m^2 u_h^2\,\beta\left(\frac{u_h}{l}\right)^{8}\right]\,.
\ee
The energy density of the black brane in the dual field theory is given in terms of the integration constant $\tilde M$ appearing in the emblackening factor \eqref{bf} as $\varepsilon = \tilde Ml^2$. This definition coincides with the usual AdS/CFT prescription where $\varepsilon=\left\langle T^{tt}\right\rangle$. In our theory,
\be\label{rho}
\varepsilon=\frac{l^2}{u_h^3}\left[1-m^2u_h^2\left(1-\frac{\beta}{7}\left(\frac{u_h}{l}\right)^8\right)\right]\,.
\ee
Moreover, we define the \textit{hydrodynamic} pressure $\mathcal{P}$ to be $\mathcal{P}\equiv \left\langle T^{xx}\right\rangle=\varepsilon/2$. Note that $\mathcal{P}$ is not identical to the \textit{thermodynamic} pressure $p$ defined in terms of the free energy density $\Omega$ as $p = -\Omega$. It is the hydrodynamic pressure which will enter the relation \eqref{ct1} between the velocity of the transverse (gapped) phonons and the rigiditiy.

\subsection{Perturbations}\label{sec:inhom}
The rigidity of the holographic solid is encoded in transverse vector perturbations which are given by 
\be\label{defb}
g_{\mu\nu}=\hat g_{\mu\nu}+h_{\mu\nu}\,,\qquad h_{ij}=\frac{1}{u^2}\left(\partial_i b_j+\partial_j b_i\right)\,,\qquad\phi^I=\hat\phi^I+\varphi^I\,
\ee
with $\partial^i b_i=\partial_i \varphi^i=0$. Together with $h_{ti}$ and $h_{ui}$ that obey similar transversality conditions this forms a set of four vector fields. Given the diffeomorphism invariance of the theory, one of the components is redundant and can be removed by a coordinate transformation leaving us with three independent vector fields. We can write these in a gauge invariant form as \cite{Baggioli:2014roa}:
\be\label{defu}
T_i\equiv u^2 \left[h_{ti}-\frac{l^2}{u^2}\partial_t\varphi_i\right]\,,\qquad U_i\equiv f(u)\left[h_{ui}-\frac{l^2}{u^2}\partial_u\varphi_i\right]\,,\qquad B_i\equiv b_i-l^2\varphi_i\,.
\ee
The equations of motion allow one to express the field $T_i$ in terms of the other two fields leaving us with only two independent equations of motion for the fields $U_i$ and $B_i$. In terms of the dimensionless quantities
\be
z=\frac{u}{u_h}\,,\quad\tilde \omega=\omega u_h\,,\quad\tilde k = ku_h\,,\quad\tilde m=mu_h\,,\quad\tilde B_i=\frac{B_i}{u_h^3}\,,\quad\tilde U_i=zU_i
\ee
the equations of motion for the Fourier components of the fields then read:
\begin{align} \label{eomU}
&\frac{1}{z^{2}}\partial_z\left[ \frac{f z^2}{V_X}  \partial_z \left(V_X \frac{\tilde U_i}{z}\right)\right]
+\left[\frac{\tilde\omega^2}{f} -\tilde k^2-2 \tilde m^2V_X\right] \frac{\tilde U_i}{z}  =  
  \frac{f' \tilde k^2}{z^2} \tilde B_i   \,,   \\[0.1cm] \label{eomB}
&z^2{\partial_z}\left(\frac{f}{z^2} \, \partial_z \,\tilde B_i \right)
+\left[\frac{\tilde\omega^2}{f} -\tilde k^2-2 \tilde m^2V_X \right] \tilde B_i 
=-2 \frac{V_X'}{V_X} \,z\,\tilde U_i \,.
\end{align}
In these expressions $f'(z)=\partial_z f(z)$ denotes a derivative with respect to $z$ while the subscript $X$ denotes the derivative with respect to $X$. In particular, $V_X'=\partial_z(\partial_XV(X(z)))$. We also note that according to the definitions \eqref{defb}, \eqref{defu} the field $B_i$ has a dimension $[B_i]=L^3$ whereas the field $U_i$ is already dimensionless. We have introduced $\tilde U_i$ since in the cases considered below this will be the field with the correct scaling dimension. Solving these equations of motion with appropriate in-falling horizon conditions will give the spectrum of the quasi-normal modes within massive gravity or, alternatively, the poles of retarded Greens functions in the dual field theory. This will be done in Sec.~\ref{sec:qnms}.

In order to compute the rigidity, it is sufficient to consider the homogeneous perturbations, i.e. to take the limit $\tilde{k} \rightarrow 0$ in the equations \eqref{eomU} and \eqref{eomB}. In particular, we allow for non-normalizable modes for $\tilde{B}_i$ but not for $\tilde{U}_i.$ Hence, the equation \eqref{eomU} admits the trivial solution $U_i=0$, while the equation \eqref{eomB} reduces to 
\be\label{spin2}
z^2\partial_z\left(\frac{f}{z^2} \partial_z \tilde{B}_i \right) + \left(\frac{\tilde\omega^2}{f} -2\tilde m^2 V_X\right) \tilde{B}_i =0 \; ,
\ee
An exact result for the Green's function can be obtained by numerically integrating \eqref{spin2}.\\
In the limit of small graviton mass $\tilde m\ll 1$ it is possible to obtain an analytical perturbative result for both the real and the imaginary parts of such correlator (see \cite{Alberte:2016xja}).

\subsection{Translational symmetry breaking pattern}\label{sec:SB}

Let us now comment on how the translational invariance is broken in this model. Specifically, whether it is possible to have a clear notion of either spontaneous or explicit breaking. As advanced in the Introduction, the only gauge invariant measure of how much the translations are broken is proportional to the graviton mass parameter\footnote{As shown in \cite{Baggioli:2014roa,Alberte:2015isw,Baggioli:2016oqk}
the model \eqref{action} and background solutions \eqref{solmetric}, \eqref{bf} can be reformulated in an equivalent way by working in the unitary gauge, which makes no reference to scalar fields getting an expectation value. Therefore the na\"ive identification of the scalar fields gradient as the order parameter seems a gauge-dependent statement.
In any gauge, however, the net physical effect is to provide a mass term for the physical spin 2 and shear modes in the gravity side.}. 
In the absence of gravity the order parameter is given by the energy density of the material (or the elastic moduli for solids \cite{Lubensky}). In the presence of gravity the order parameter gives rise to the  plasma mass terms for the graviton, see \cite{Alberte:2015isw}.

From the equations of motion \eqref{eomU}, \eqref{eomB}, and \eqref{spin2} we see that the mass of the dynamical fields is controlled by the quantity
\be
M^2(z)\equiv \tilde m^2V_X=m^2u_h^2\left(1+5\beta \left(\frac{u_h}{l}\right)^{8}z^8\right)
\ee
that is a function, dependent on the radial coordinate.  In particular, one can define the ultraviolet (UV) and infrared (IR) masses as the values of $M^2(z)$ at $z=0$ and $z=1$ respectively:
\be\label{uvir}
m_{UV}^2=m^2u_h^2\,,\qquad m_{IR}^2 =m^2u_h^2\left(1+5\beta \left(\frac{u_h}{l}\right)^{8}\right) \,.
\ee
Their ratio,
\be
\frac{m_{UV}^2}{m_{IR}^2}=\frac{1}{1+5\beta \left(\frac{u_h}{l}\right)^{8}}\,,
\ee
is clearly  controlled solely by the parameter $\beta$ in the bulk Lagrangian. 

In the CFT language, the fact that the natural order parameter for the translational symmetry breaking in the bulk is a function of the radial coordinate translates as the fact that the translations are broken by different amounts at different energy scales. The effect could be similar to a chain of different `super-lattices' on top of underlying `sub-lattices', for instance. For the practical purposes of this work we will not enter into whether this is a correct interpretation. Instead we simply note that for fairly general choices of $V(X)$ one can arrange a simpler situation where the dynamics are controlled only by two relevant scales. For $V(X)$ such that $M(z)$ is approximately constant during most of the flow and then suddenly grows towards the IR, 
the relevant scales are the asymptotic values of $M(z)$---the IR and UV masses $m_{IR}$ and $m_{UV}$. This identification is of course imperfect, but by taking large enough $\beta$ and $N$ in \eqref{bench} we can get sufficiently close to it.

The bottom line of this discussion is that for these models it is natural to expect that there are two different sources of the translational symmetry breaking, controlled by two different mass scales. When this is the case, we are under the conditions of having light gapped phonons provided that the ratio $m_{UV}/m_{IR}$ is small enough. 

Let us argue now how large should the phonon mass gap be in relation to the parameter controlling the amount of the explicit versus spontaneous breaking, $m_{UV}/m_{IR}$.
We recall that in the case of the chiral symmetry breaking in QCD the Goldstone boson mass squared is proportional to the quark masses (the source of the explicit breaking)
$m_\pi^2 \sim m_q$. By analogy 
we expect that the phonon mass, $\omega_0$, should be proportional to the square-root of the explicit breaking source.
The numerical results presented in Sec.~\ref{sec:qnms} provide a strong indication that this intuition is correct. On the right panel of Fig.~\ref{fig:qnms_beta} we see that the ratio $\omega_0/\Lambda\approx |\omega_1|/|\omega_2|$ quantifying the separation between the two lowest quasi-normal modes does indeed scale as $\sqrt{m_{UV}/m_{IR}}$.

\section{Transverse quasi-normal modes}\label{sec:qnms}

 In this section we compute the quasi-normal modes (QNMs) of the transverse sector \eqref{eomU} and \eqref{eomB} to show the effect of $\beta$ on the spectrum, to identify the phonon-like excitations and to check the expected relationship \eqref{sound_rel} between the velocity of the transverse phonon and the elasticity. A detailed discussion on how to obtain these modes and on the numerics can be found in Appendix~\ref{app:numerics}. In the presence of $\beta$ the model has two independent dimensionless parameters. We have used the scaling symetries of the equations to set $l=u_h=1$ and therefore consider $T/m$ and $\beta$ as the independent parameters describing our system.\footnote{The scaling symmetries of the equations show that in terms of the CFT quantities the parameters $m/T$ and  $\beta\, u_h^8$ are scaling invariants. For convenience, due to the small temperatures considered, we plot against $\beta$ in the figures. In all our computations we have set $u_h=1$, therefore when we refer to  $\beta\gg1$ it should be understood as $\beta\,u_h^8\gg 1$ or  $ \beta/ s^4\gg1$.}

 \subsection{The quasi-normal mode spectrum}
 We first discuss the QNMs at zero momentum and compare them to the known behaviour for $\beta=0$ (see \cite{Davison:2014lua}).
\begin{figure}
\center
\includegraphics[width=0.48\textwidth]{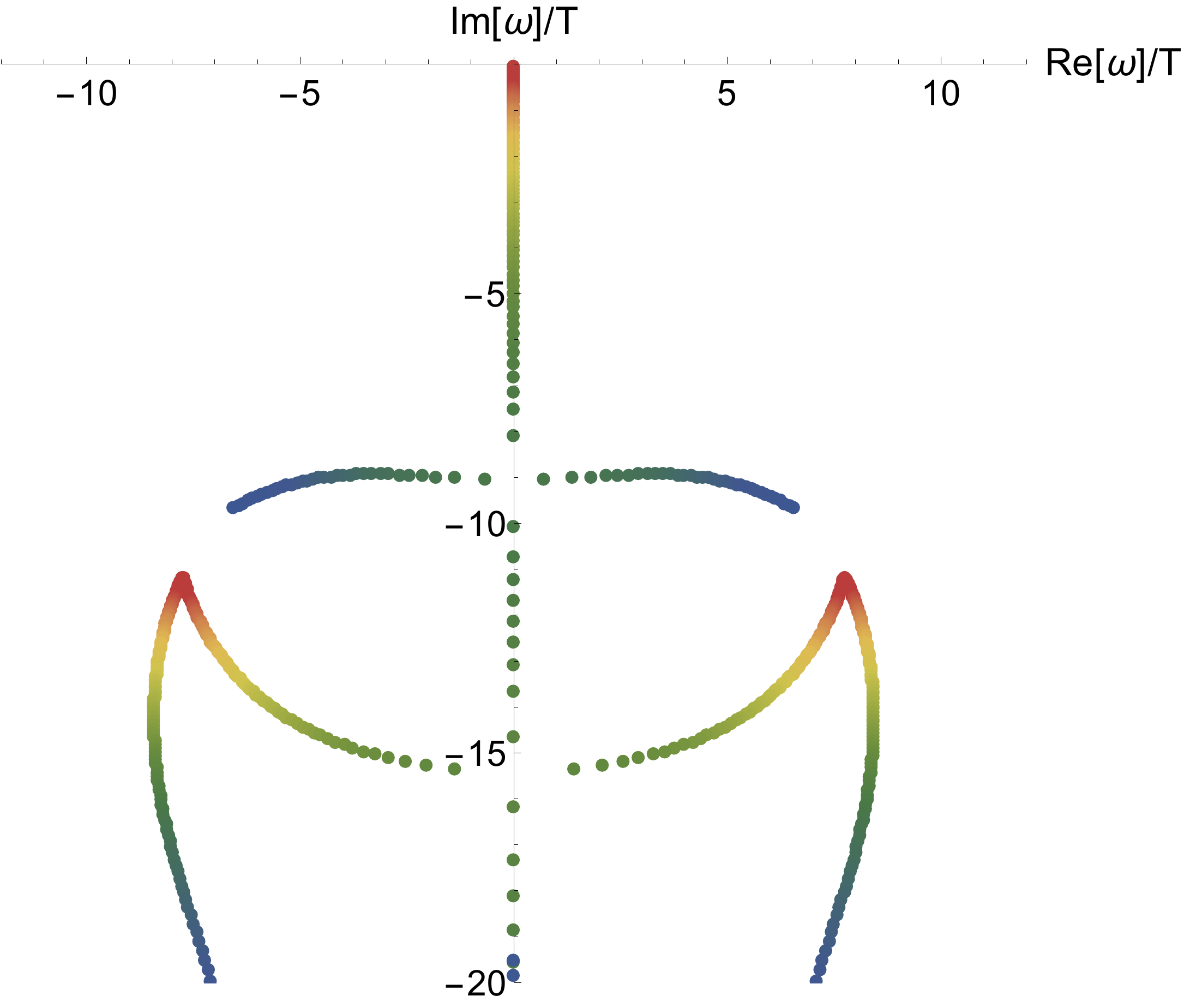}
\includegraphics[width=0.48\textwidth]{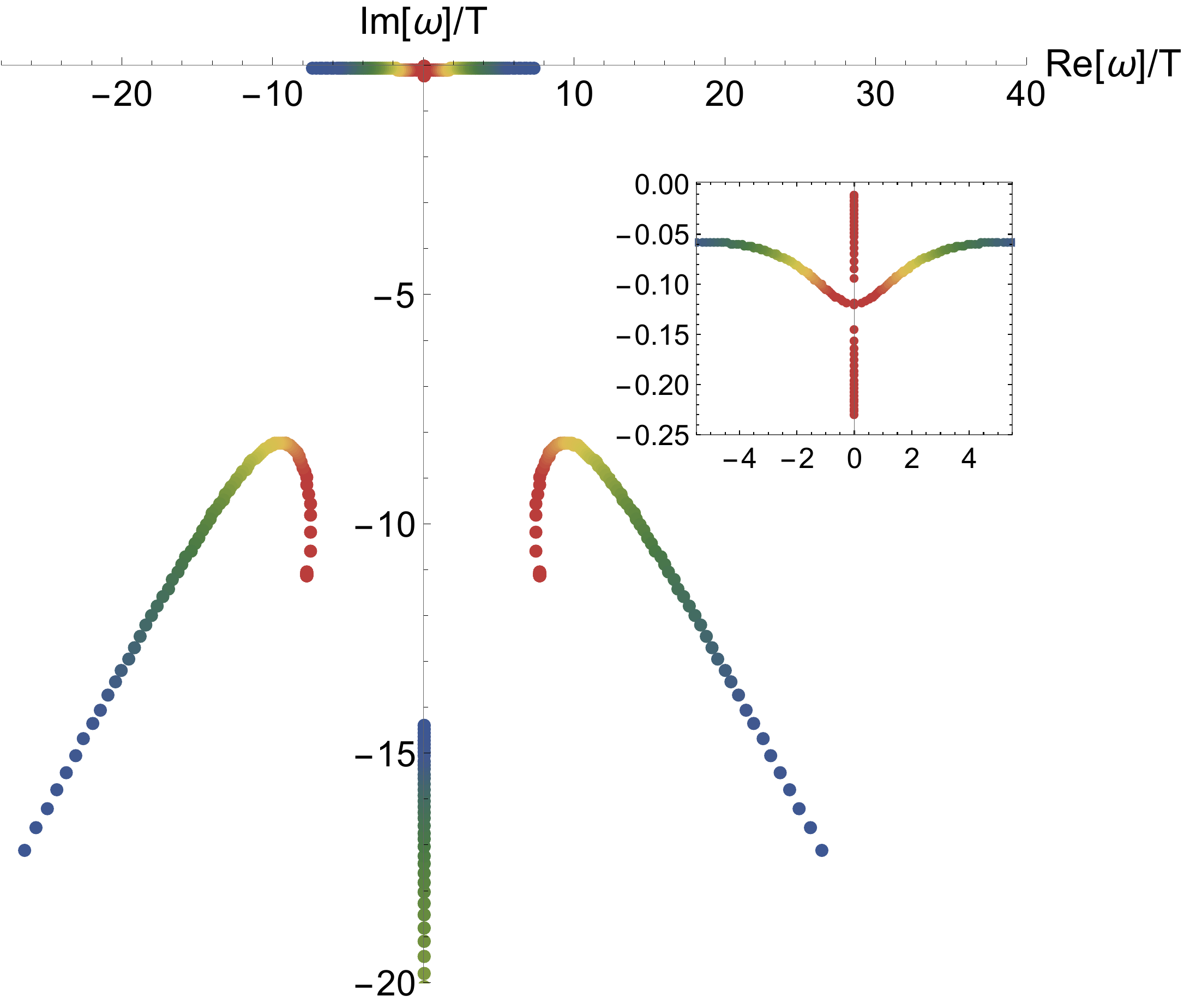}
\caption{The real and imaginary parts of the lowest quasi-normal modes at $k=0$. \textbf{Left:} $\beta=0 $ and $ T/m\in[0.10, 1.07]$ (blue--red). \textbf{Right:}  $\beta=60$ and $T/m\in [0.16, 68.97 ]$ (blue--red).}
\label{fig:betak0}
\end{figure}
 The spectrum of the lowest QNMs at zero momentum as a function of $T/m$ for $\beta=0$ and $\beta=60$ \footnote{The results at this particular value of $\beta$ are rather generic for any finite value of $\beta$. Possible differences appearing at other values are discussed in the text.} is shown in Fig.~\ref{fig:betak0}. In both cases we see that at high enough temperatures $T/m$ (red) there is a single purely imaginary mode whose absolute value is small compared to that of the next lowest mode. By decreasing $T/m$ below a critical value the two lowest purely imaginary modes collide developing a non-zero real part in the frequency (henceforth we refer to this as the \emph{collision point}), in agreement with \cite{Davison:2014lua} for $\beta=0$.\footnote{We note that our convention for $m^2$ differs from that of \cite{Davison:2014lua} as follows: $m^2_{\text{there}}=2m^2_{\text{here}}$.}

\begin{figure}
\begin{center}
\includegraphics[width=0.47\textwidth]{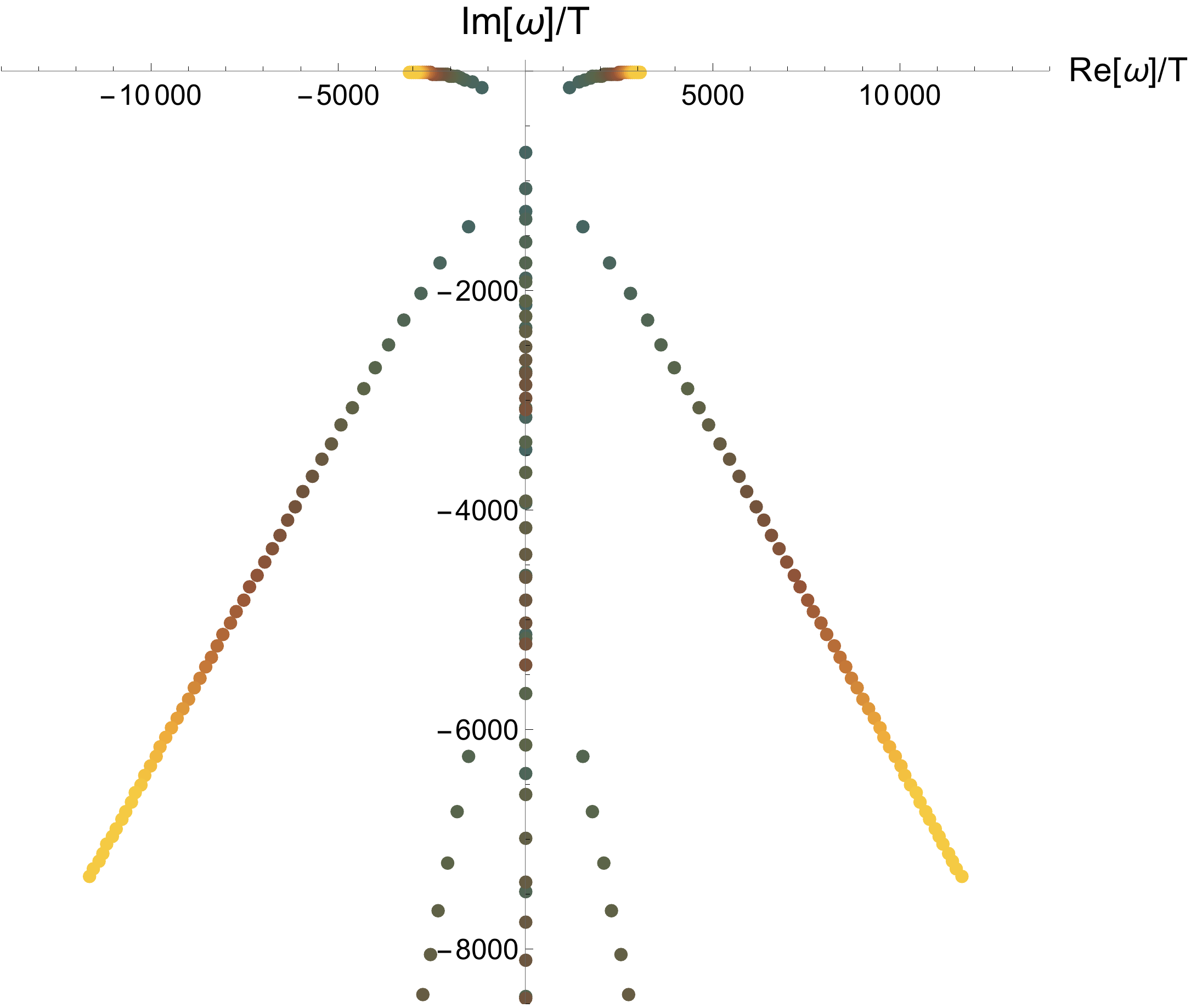}
$\,\,\,\,\,\,\,\,\,$
\includegraphics[width=0.47\textwidth]{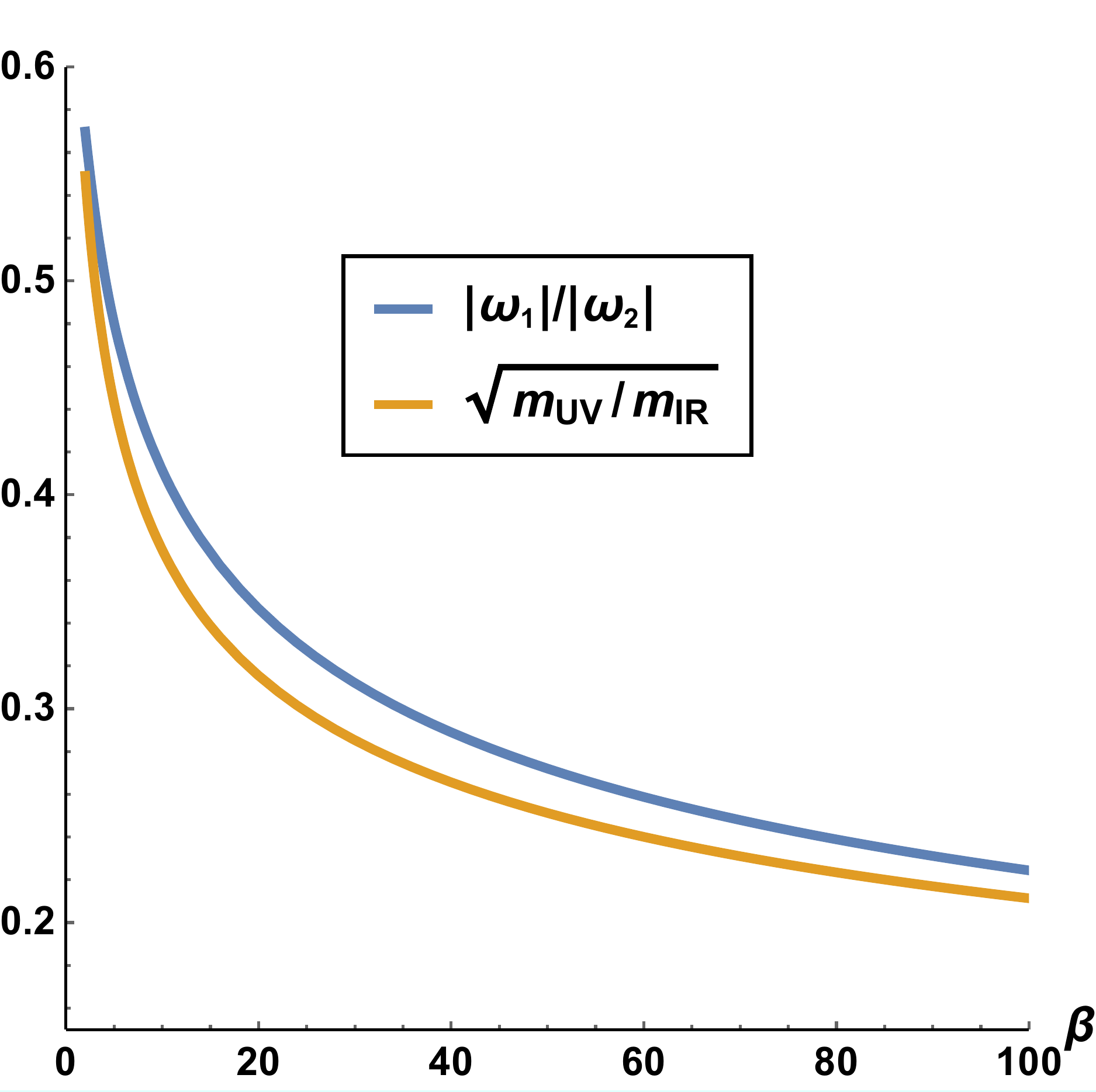}
  \caption{\label{fig:qnms_beta}\textbf{Left:} QNMs at k=0 for $T/m=0.001$ and $\beta\in [1,100]$ (green--yellow). A qualitatively similar behaviour is found also for other values of $T/m$. \textbf{Right:} Comparison between the ratio of the absolute values of the frequencies of the first and the second lowest QNMs shown on the left panel (blue) and the mass ratio $\sqrt{m_{UV}/ m_{IR}}$ (yellow).}
\end{center}
\end{figure}

Let us examine the effects of a non-vanishing $\beta$. The evolution of the QNMs at zero momentum with increasing $\beta$ at fixed temperature $T/m=0.001$ is depicted in the left panel of Fig.~\ref{fig:qnms_beta}. We see that by increasing $\beta$ the value of the real part of the lowest mode increases while the absolute value of its imaginary part decreases. This implies that for increasing values of $\beta$ the lowest QNM becomes less and less damped thus indicating the presence of a long lived excitation. This is in fact already evident by comparing the two panels in Fig.~\ref{fig:betak0}. In particular, for $\beta=0$ the imaginary part at the collision point is $\omega/T\sim -10i$. Hence, at the moment when the frequency of the lowest quasi-normal mode develops a non-zero real part its decay rate $\Gamma\sim T$, so that there is no long-lived excitation in the spectrum. On the other hand, for $\beta=60$ we find $\omega/T\sim -0.1i$ corresponding to a parametrically much smaller decay rate. Another important effect due to increasing of $\beta$ is that the rest of the QNM spectrum moves away from the origin, thus generating a stronger hierarchy between the modes. Hence, we see that for large values of $\beta$ the momentum transport in the transverse sector is governed by two long lived gapped modes, well-separated from the rest of the QNM tower. We identify these with the pseudo-Goldstone bosons due to the spontaneous breaking of translational invariance.

To better illustrate this point, we can specifically look at the separation of scales between the first and second QNMs. In the right panel of Fig.~\ref{fig:qnms_beta} we plot the ratio between the absolute values of the frequencies of the first and second QNMs, as well as the ratio of the graviton mass terms in the UV and in the IR, as defined in \eqref{uvir}. Even if the two ratios are not in a perfect proportion, we can see a clear correlation in that they both decrease monotonically in $\beta$. This justifies that we can associate the origin of the phonon mass gap with the presence of a small but nonzero $m_{UV}$ (as compared to $m_{IR}$), as argued in Sec.~\ref{sec:SB}

\begin{figure}[htp]
\begin{center}
\includegraphics[width=0.48\textwidth]{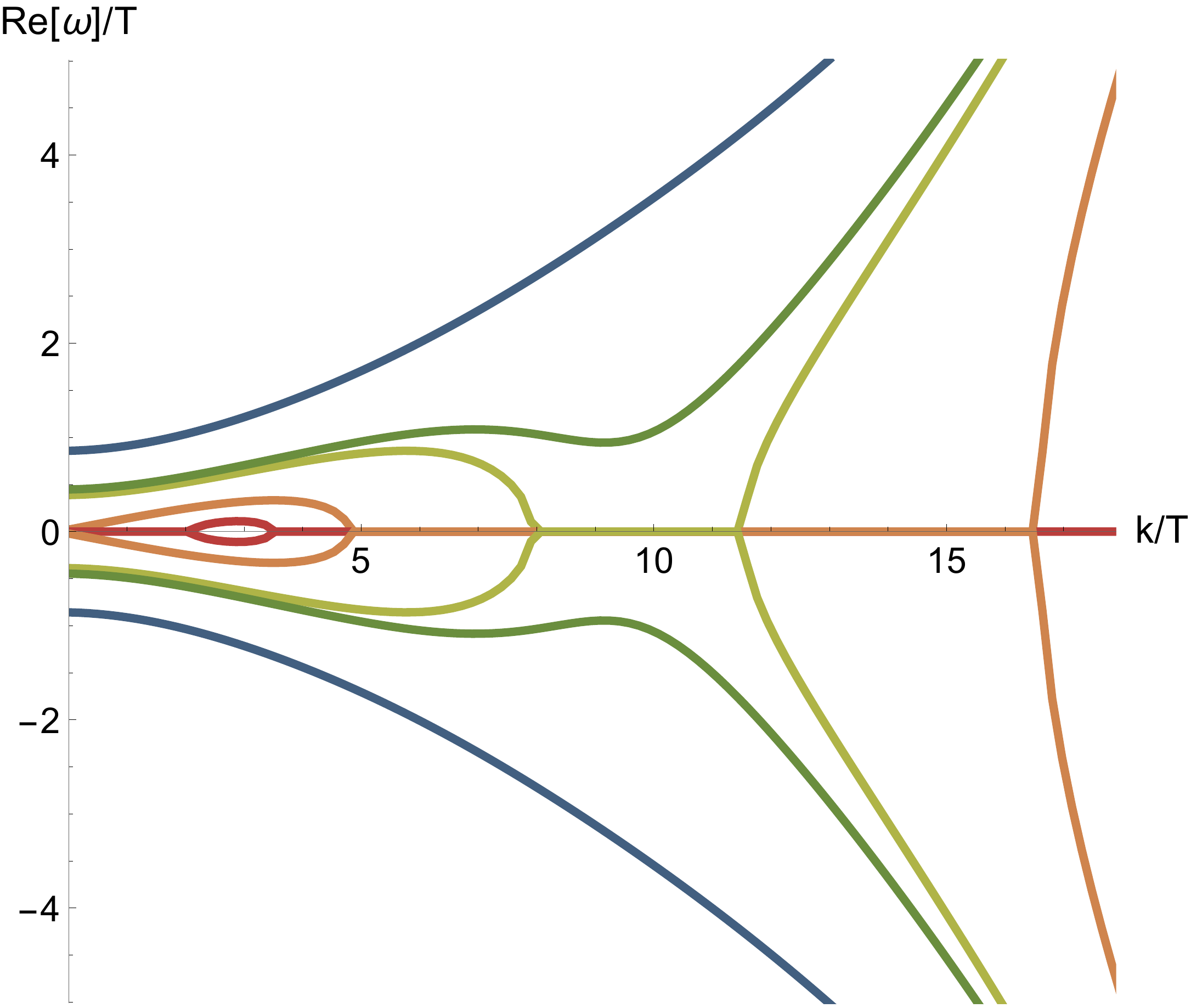}
\includegraphics[width=0.48\textwidth]{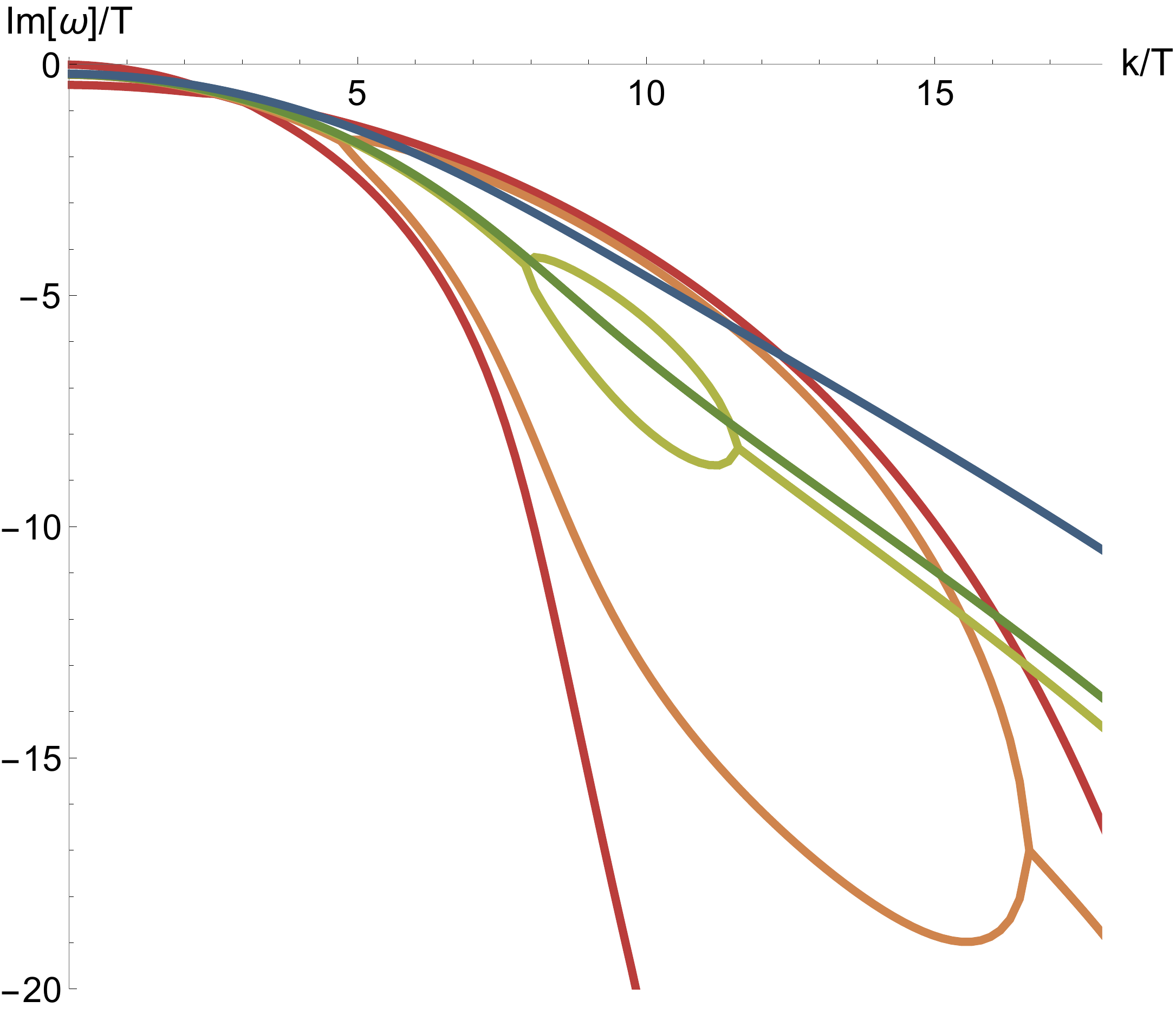}
  \caption{\label{fig:modeswithk}The real (\textbf{left}) and imaginary (\textbf{right}) parts of the frequencies of the two lowest QNMs at $\beta=20$ for $T/m\in[2.89,68.27]$ (blue--red).}

\end{center}
\end{figure}

\begin{figure}[htp]
\begin{center}
\includegraphics[width=0.48\textwidth]{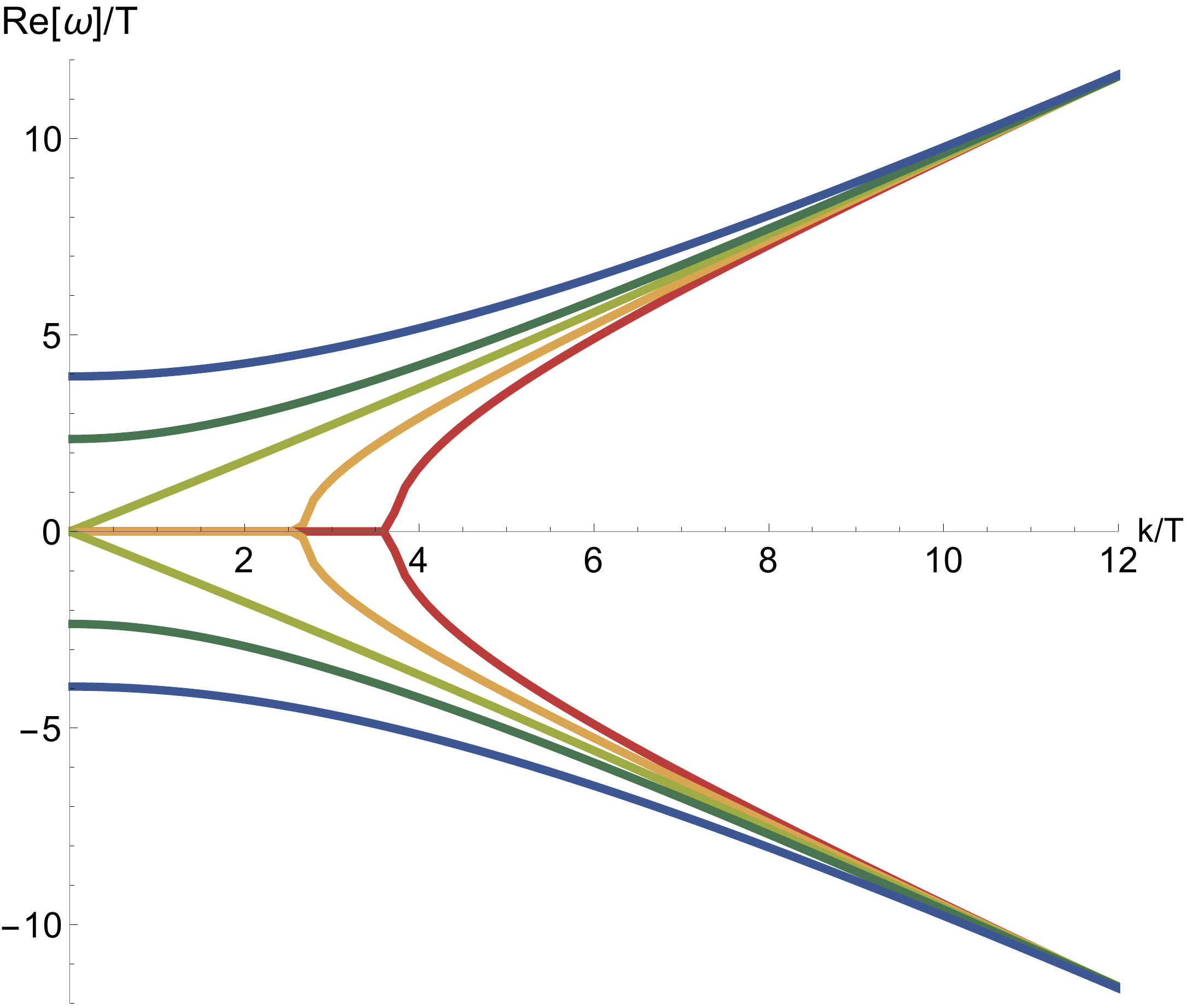}
\includegraphics[width=0.48\textwidth]{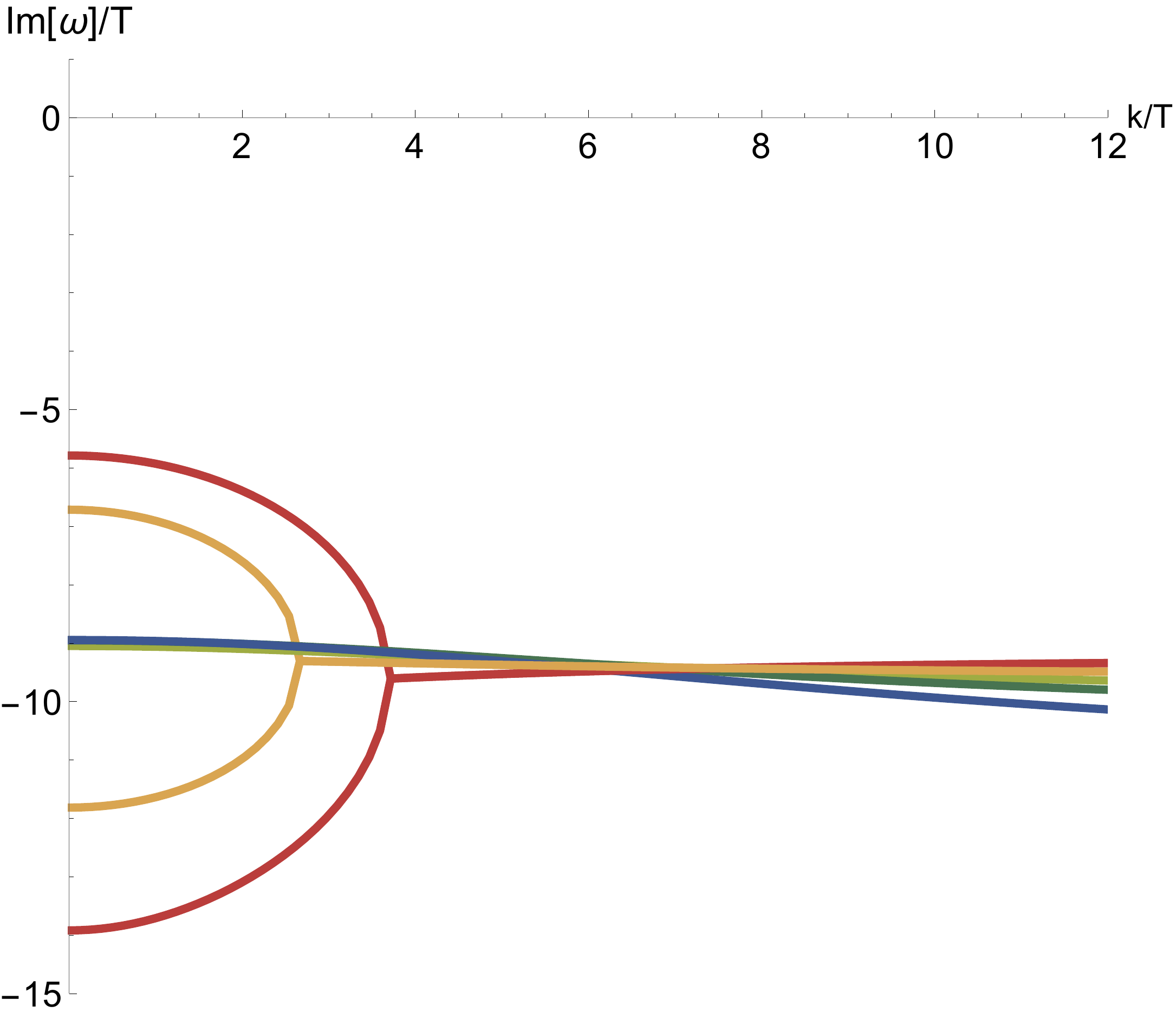}
  \caption{\label{fig:withkbeta0}The real (\textbf{left}) and imaginary (\textbf{right}) parts of the frequencies of the two lowest QNMs at $\beta=0$ for $T/m\in[0.14, 0.16]$ (blue--red).}
\end{center}
\end{figure}

Finally, we study the effect of $\beta$ on the lowest QNMs at finite momentum. In Fig.~\ref{fig:modeswithk} we show the real and imaginary parts of the frequencies of the two lowest QNMs as a function of momentum at $\beta=20$. The different colors in Fig.~\ref{fig:modeswithk} correspond to different values of $T/m$, changing from blue to red as the temperature increases.  Let us first discuss the behaviour of the real part distinguishing three different energy regimes. At high enough momentum with $k/T\gg1$, all the different temperature modes converge to the same slope, as one would expect in this energy regime. On the other hand, at very low energies, for $k/T\ll1$, we find three qualitatively different behaviors, depending on the temperature. First, there exists a temperature $T/m$ at which the mode shows no gap in the real part; this is the collision point (orange line). At this temperature the dispersion relation of the lowest mode looks like $\text{Re}[\omega]\sim c\,k$. Let us remark that the temperature of the collision point increases with $\beta$.
Above this temperature the mode is purely diffusive (red line) while below it the mode is gapped. In the latter regime the  dispersion relation takes the form $\text{Re}[\omega] \sim a + v\,k^2$ (green and blue lines). These two regimes---the far UV and the deep IR---show a  behaviour qualitatively similar to the case $\beta=0$, shown in Fig.~\ref{fig:withkbeta0}, see \cite{Jimenez-Alba:2014iia,Grozdanov:2016vgg, Hofman:2017vwr} for further discussions on this behavior in holography and \cite{Stephanov:2014dma} in kinetic theory. The effect of $\beta$ in these regimes is only quantitative: it alters the specific values of $k/T$ at which the qualitative behavior of the modes change, as well as the specific values of $\omega/T$. It is in the intermediate energy regime where $\beta\neq 0$ induces a qualitatively new behavior, as can be seen in Fig.~\ref{fig:modeswithk}. In particular, for some finite values of $\beta$ the real part of the modes decreases with increasing finite momenta, showing a dispersion relation that resembles that of rotons (green line). Eventually, for high enough temperatures the two modes recollapse and become purely imaginary again at (yellow, orange and red lines in Fig.~\ref{fig:modeswithk}). This phenomenon is absent for $\beta=0$ as shown in Fig.~\ref{fig:withkbeta0}.

The imaginary part of the frequency behaves ``complementarily'' to the real part in the following sense. When the modes are propagating, \emph{i.e.} the real parts are non-zero, the imaginary parts are identical and both modes have the same decay rate. On the contrary, when the modes become purely imaginary, they decouple and show different decay rates. Similar behavior has been previously found in holography \cite{Gursoy:2013zxa, Janik:2016btb}. \\

\subsection{Transverse velocity and elasticity}

As  explained in previous sections, we argue that at least in some regime of the parameter space the lowest collective excitations behave as transverse gapped phonons. 
In order to support this claim we compare the velocity of the lowest QNM to the the standard expression for the transverse sound speed in solid materials given in \eqref{sound_rel}.

The shear modulus $\mu$ in \eqref{sound_rel} can be obtained \cite{Alberte:2015isw,Alberte:2016xja} as the real part of the retarded Green's function of the transverse traceless metric perturbations at zero momentum and frequency \eqref{spin2}. This leads to the following expression for the sound speed of the transverse excitations:
\be\label{ct2}
c_T^2=-\frac{3}{2L}\frac{1}{\varepsilon+\mathcal P}\,\lim_{z,\tilde\omega\to 0}\text{Re }\frac{1}{6}\frac{\tilde{B}_i'''(z)}{\tilde{B}_i(z)}\left(\frac{l}{u_h}\right)^3\,,
\ee
 where the energy density $\varepsilon$ is given by the black brane energy density  \eqref{rho} and $\mathcal P=\varepsilon/2$ is the hydrodynamic pressure appearing in the stress-energy tensor at the boundary. We compute this by solving \eqref{spin2} numerically for zero momentum. An analytic expression valid in the $m/T\to 0$ limit was previously obtained in \cite{Alberte:2016xja}.\\

On the other hand, the velocity of the collective excitations can be obtained from the real part of the frequency of the lowest QNM at finite, low momentum $k/T \ll1$. As mentioned in the previous section, at temperatures below the collision point and for small momenta, the real part of the lowest mode can be fitted to
\be\label{fit}
\omega=\sqrt{a+v^2\,k^2}\,,
\ee
with $a=a(T/m,\beta)$ vanishing at the collision point and giving rise to a linear behavior. To be more precise, we have fitted the data to

\begin{equation}\label{fit2}
(\text{Re}[\omega])^2= a_1+ a_2\,k^{a_3} \, ,
\end{equation}
with $a_3=2\pm10^{-4}$ for the parameters investigated in this paper. 

\begin{figure}[htp]
\begin{center}
\includegraphics[width=0.47\textwidth]{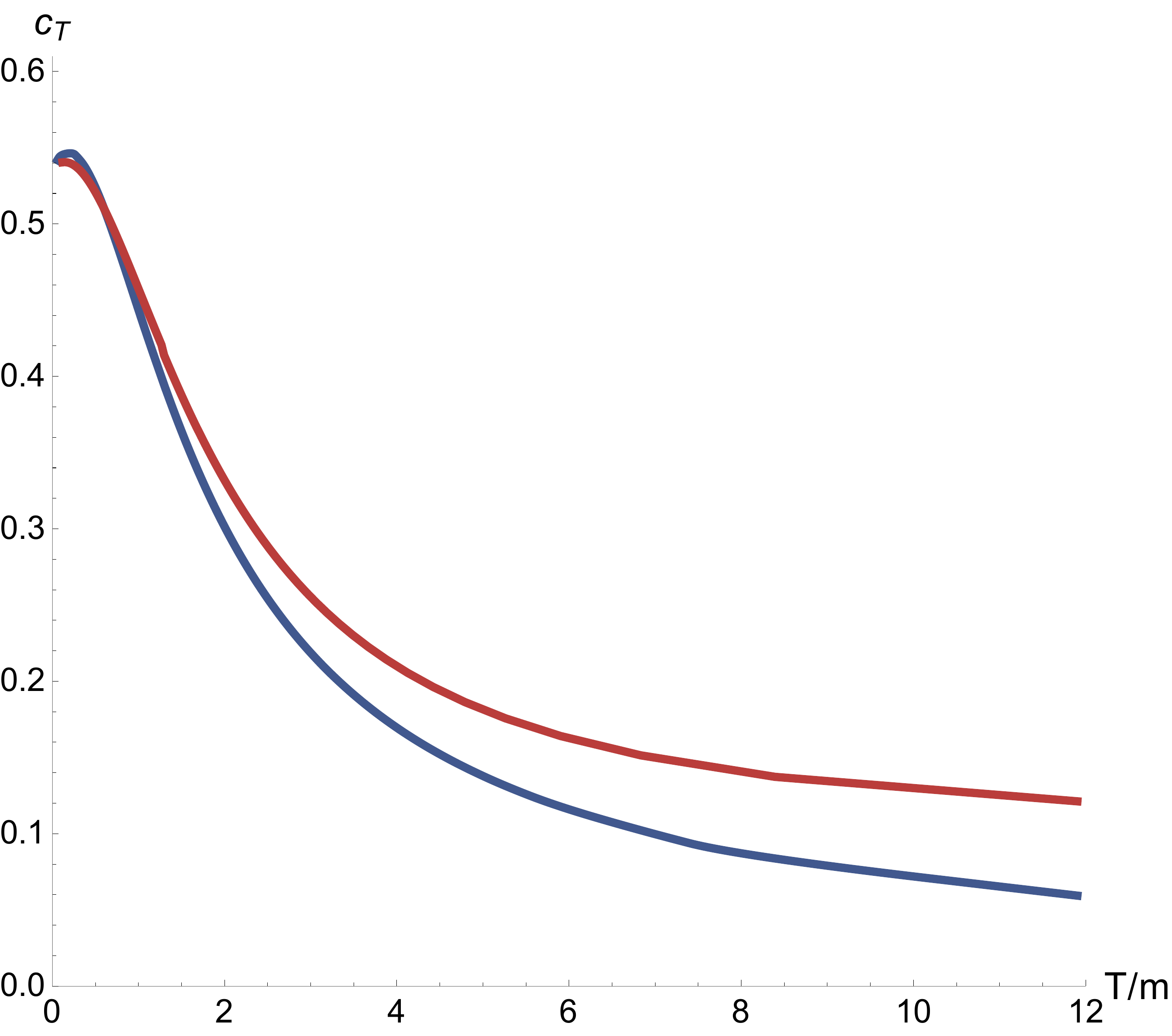}
\includegraphics[width=0.47\textwidth]{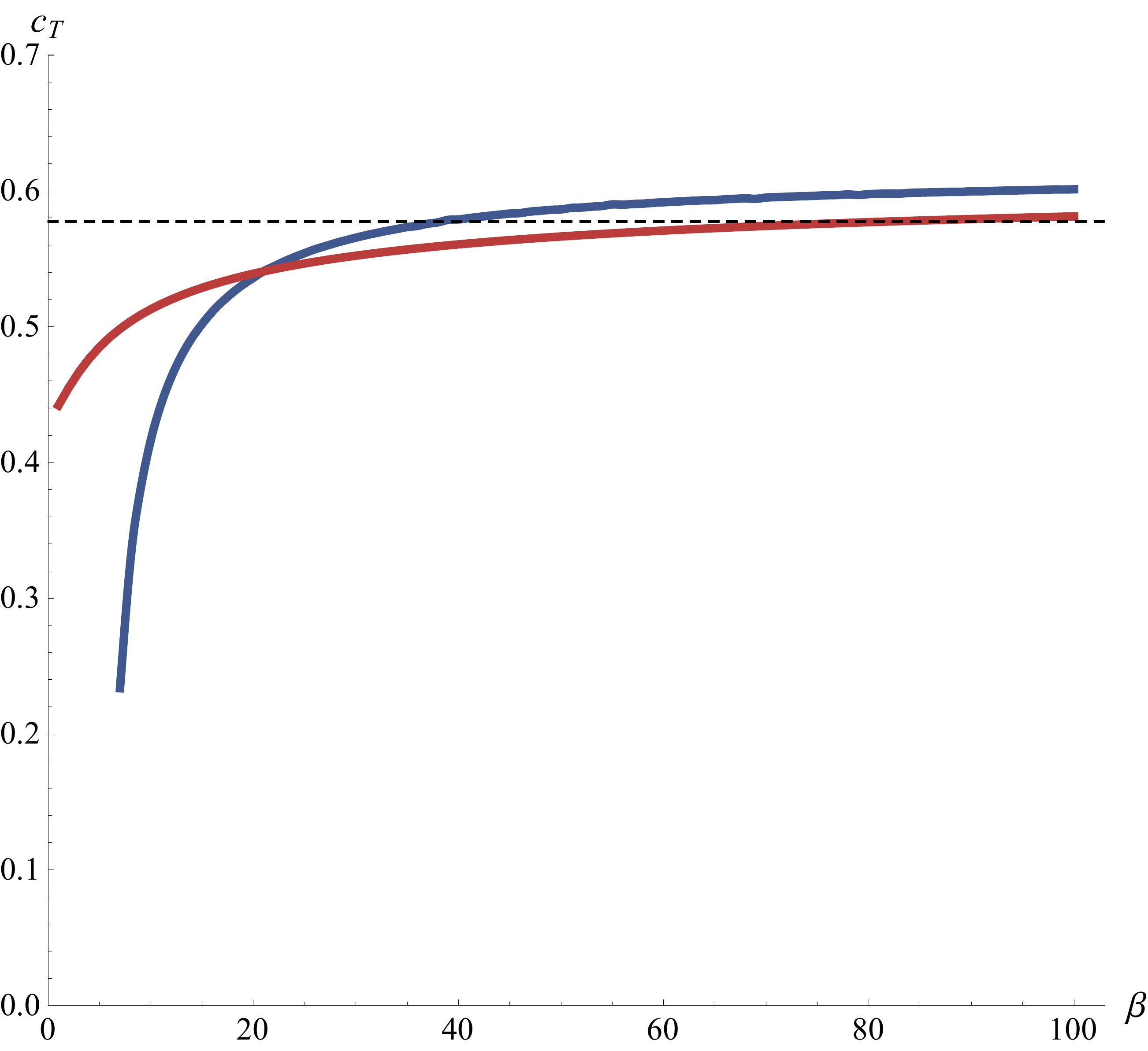}
 \caption{\label{fig:velocity}Comparison between the velocity obtained from the QNMs (red line) to the velocity given in Eq.~\eqref{ct2} (blue line). \textbf{Left:} The velocity as a function of temperature for $\beta=20$. \textbf{Right:} The velocity as a function of $\beta$ for $T/m=0.05$. (As a cross-check let us mention that the velocities evaluated at $T/m=0.001$ coincide with the results shown here.)}
\end{center}
\end{figure}
In Fig.~\ref{fig:velocity} we show the comparison between the sound speed determined from the quasi-normal mode spectrum by the fit \eqref{fit2} and the expression \eqref{ct2}. On the left panel we show the comparison for fixed $\beta=20$ as a function of temperature $T/m$. We see that the agreement between the speed of propagation extracted from the spectrum of QNMs and the velocity computed by the standard EFT expressions \eqref{sound_rel} and \eqref{ct2} is not exact and seemingly improves at low temperatures. This is however a coincidence and is not the case for other values of $\beta$. The actual reason behind the disagreement is the presence of the small but finite explicit breaking---it is indeed particularly pronounced at small values of $\beta$ where the explicit breaking is large. This is confirmed by what we see in the right panel of Fig.~\ref{fig:velocity}.
We further observe that as $T/m\to 0$ the sound speed approaches some maximal value and continuously decreases with increasing temperature. In other words, it shows a melting-like behaviour. At high temperatures the system is a fluid in which there are no transverse sound waves and thus $c_T^2=0$. At intermediate temperatures it is a viscoelastic material with finite viscosity $\eta$ and a non-zero shear modulus $\mu$. In distinction from melting that takes place in ordinary solids, the transition to the fluid phase is continuous in these materials.

On the right panel of Fig.~\ref{fig:velocity} we plot the velocity as a function of $\beta$ at fixed $T/m=0.05$.  From here we see that the maximal speed of sound reached at low temperatures flattens out in the large $\beta$ limit. Despite the general similarity, we do see a slight disagreement between the velocity obtained from the QNMs and the EFT expectations, persistent even in the regime of low temperatures and high values of $\beta$. We believe that this is a combined effect due to a non-zero mass gap and a non-zero decay rate of our collective excitations. Neither of these effects is taken into account in \eqref{sound_rel}. It would be interesting to gain a quantitative understanding of this deviation. We leave it for future work. Another interesting feature that Fig.~\ref{fig:velocity} seems to suggest is that the sound speed might saturate in the large $\beta$ limit to some constant value that is close to $c_T^2=1/3$. That this might be a universal feature for the transverse sound speed in conformal solids is also suggested in \cite{Esposito:2017qpj} and deserves further investigation.
\\

\section{Discussion}\label{sec:discussion}

In this work we have shown how the elastic properties of black holes are encoded  in certain gravity theories that break translational invariance and therefore allow for momentum dissipation. We have shown that the standard relations from the elasticity theory are realized in these solutions to a very good accuracy, and that a clear identification of gapped but {\em light} transverse phonons is built in the framework. 

By the AdS/CFT correspondence, these black holes can be interpreted as strongly coupled theories that describe solid materials in a critical regime, with the phonon-like excitations being part of the dynamical spectrum of the theory.

These elastic phenomena are realized in BHs within the class of holographic massive gravity theories introduced in \cite{Baggioli:2014roa}. Our main result is the following: In a certain (expected) parameter regime of these theories, the speed of propagation of the transverse phonons qualitatively coincides with the speed computed from the elastic rigidity modulus (as introduced in \cite{Alberte:2015isw}), see Fig.~\ref{fig:velocity}. We also discuss how to understand the small mass gap of the transverse phonons from the symmetry breaking pattern, see the right panel of Fig.~\ref{fig:qnms_beta}.

These results give explicit confirmation that the lowest modes in the shar channel can be identified as transverse phonons and therefore that the planar black brane solutions of these massive gravity theories do behave as solids, as previously advocated in  \cite{Baggioli:2014roa,Alberte:2015isw,Alberte:2016xja}.  The clear relation between the phonon mass gap and the source of explicit breaking also backs up the proposal of \cite{Baggioli:2016oqk} that these gravity duals can be used to model some aspects of disorder.
Our results open up a number of questions.

 By construction our models do not allow to switch off completely the phonon mass gap in the low temperature regime. In particular, it would be interesting to find a model realizing exactly gapless phonons. Our results suggest that the limit that could generate this is to take $m_{UV}\to0$ while keeping $m_{IR}$ finite. In this limit, the shape of the non-canonical kinetic terms for the scalar  becomes apparently singular (\emph{i.e.} $V'(X)\to0$ near the AdS boundary). We leave for future investigation whether this kind of limit is actually well defined and whether it  does give rise to gapless phonons.

     A natural continuation of our work would be to analyze the longitudinal sector of the excitations. This provides a further check of the emerging picture since the longitudinal sound speed also has to relate to the bulk and shear elastic moduli according to \eqref{speeds}.

     Another interesting point for the future is to see whether there is a BH transition at high temperature corresponding to its `melting' into a fluid. We already see some evidence for that due to the transverse phonon becoming eventually over-damped and showing a diffusive behaviour at high enough temperatures. Moreover, by increasing $T/m$ both the transverse speed of sound and the elastic modulus continuously decrease and tend to the fluid limit ($c_T=\mu=0$). This may be linked to the fact that these BHs have simultaneously elasticity and viscosity ({\em i.e.} viscoelasticity) \cite{Alberte:2016xja}. See also \cite{Kachru:2009xf,Anninos:2013mfa} for other related work in holography.

     One could also add a non-zero charge density to our setup and study in more detail the proposal of \cite{Delacretaz:2016ivq} in connection with the transport properties of bad metals. The optical conductivity for these models has already been computed in \cite{Baggioli:2014roa}; it exhibits a pinned response showing the presence of both explicit and spontaneous breaking. As a further confirmation of the present picture, the DC conductivity shows a metal--insulator transition which is typical of pinned systems (\textit{e.g.} the pinned charge density waves).\\ Additionally, it would be important to check for signatures of the presence of the phonons in the DC transport properties and thermodynamical observables, as attempted in \cite{Baggioli:2015gsa}.

     A proper and complete hydrodynamic analysis of these holographic systems with translational symmetry breaking is still lacking. It would be extremely useful to build such a framework in order to study in a more rigorous and systematic way the elastic moduli and the viscosities in terms of Kubo formulas. This sheds light on the several still present puzzles related to those physical observables.

Let us end by emphasizing one important aspect of our results. Effective Field Theory description of Solids \cite{Leutwyler:1996er,Dubovsky:2011sj,Nicolis:2013lma} is a powerful way to describe phonons in solid materials that seems to be completely orthogonal to the holographic models used here. In fact, we have used quite a lot of the intuition obtained from this EFT. A natural question then arises: is there any solid EFT that describes the properties of our elastic black holes? Our results are very encouraging although this is not obvious \emph{a priori}. The reason is simple: in order for the  EFT description to be a good approximation, there needs to be a separation of scales. In the standard (weakly coupled) solids a large separation of scales does indeed exist between the lattice spacing and the low energy phonon modes. In the holographic solids (the CFT duals of the AdS BHs studied here), however, there is no separation of scales because by construction the field theory is scale-invariant, \emph{i.e.} there is a continuum of CFT modes at $T=0$. Our results imply that it is sufficient to have a separation of scales in the QNM spectum---\emph{i.e.} that there is one QNM much lighter than the remaining ones. Hence, EFT-like relations may still hold even in conformal field theories. We find this an interesting output of our work. \\
We plan to return to these issues in the near future.

\section*{\sc Note added}

While this article was under the publication review process, a new work by the same authors \cite{Alberte:2017oqx} appeared. This work confirms the picture proposed and concludes that it is indeed possible to realize exactly gapless transverse phonons as suggested above.

\section*{Acknowledgments}

We thank Andrea Amoretti, Tomas Andrade, Aron Beekman, Angelo Esposito,  Daniel Fernandez, Blaise Gouteraux, Sean Hartnoll, Matti Jarvinen, Keun-Young Kim, Elias Kiritsis,  Alexander Krikun, Julian Leiber, Weijia Li, Yi Ling, Matthew Lippert, Andy Lucas,  Nicodemo Magnoli, Rene Meyer, Daniele Musso,  Philip Philips,  Napat Poovuttikul, Sang-Jin Sin, Saso Grozdanov and Jan Zaanen  for useful discussions and comments about this work and the topics considered.

We are particularly grateful to Matthew Lippert, Niko Jokela, Marti Jarvinen, Tomas Andrade, Alexander Krikun and Napat Poovuttikul for sharing with us informations about their unpublished works.

OP acknowledges support by the Spanish Ministry MEC under grant FPA2014-55613-P and the Severo Ochoa excellence program of MINECO (grant SO-2012-0234,  SEV-2016-0588), as well as by the Generalitat de Catalunya under grant 2014-SGR-1450. MB is supported in part by the Advanced ERC grant SM-grav, No 669288. MB would like to thank the Asia Pacific Center for Theoretical Physics, Hanyang University and Gwangju Institute of Science and Technology for the hospitality during the completion of this work. MB would also like to thank the Asia Pacific Center for Theoretical Physics for the large quantity of coffee needed to complete this manuscript. AJ acknowledges financial support by Deutsche Forschungsgemeinschaft (DFG) GRK 1523/2.

\appendix
\section{Solid EFT and elasticity}\label{app:eft}
In this section we derive the expressions \eqref{sound_rel} for the transverse and longitudinal sound speeds in an elastic medium with non-negligible pressure. We use the standard EFT methodology for describing solid materials \cite{Leutwyler:1993gf, Leutwyler:1996er, Dubovsky:2011sj,Nicolis:2015sra,Nicolis:2013lma} in flat space-time. In $3+1$ dimensions, the low-energy behavior of solids can be described by using three scalar fields $\phi^I(x^i,t)$, corresponding to its comoving coordinates. The effective field theory of the long wavelength excitations in a solid is then given by the scalar field action
\be\label{lagr}
S=-\int d^{3+1}x\,F(X,Y,Z)\,
\ee
where
\be
X  \equiv \tr[\I^{IJ}]\,,\qquad Y\equiv\tr[\mathcal I^{IJ}\mathcal I^{JK}]\,,\qquad Z\equiv\text{det} \left[\mathcal I^{IJ}\right]\,,\qquad \mathcal I^{IJ}\equiv\d_\mu \phi^I \d^\mu \phi^J\,.
\ee
The indices $I,J=\{x,y,z\}$ are contracted with $\delta_{IJ}$. In this section the spacetime metric is set to the flat Minkowski metric $\eta_{\mu\nu}=\text{diag}(-1,1,1,1)$.

Let us consider an isotropic and homogeneous background field configuration
\be\label{vev}
\hat \phi^I = x^i \delta^I_i\,,
\ee
corresponding to the equilibrium configuration of the solid. On this background, the full stress tensor of the scalars
takes a perfect fluid form $T^\mu_{\ \nu}=\text{diag}(-\varepsilon,p,p,p)$ with energy density and pressure identified as
\be\label{rhop}
\varepsilon= F \, , \qquad \quad  p = - F +\frac23 \left(X F_X +2YF_Y+3ZF_Z\right)~ ,
\ee
all quantities evaluated on the background \eqref{vev}.

It is then easy to see how the functional form of $F(X,Y,Z)$ relates to the linear elastic response parameters $\mu$ and $\kappa$ defined in \eqref{sigma1}. These coefficients are a measure of the strength with which the solid opposes a certain deformation, described in terms of the displacement vector $u_i$. In particular, the stress tensor due to a traceless shear deformation is given in terms of the shear modulus $\mu$ in Eq.~\eqref{sigma2}.

In the EFT setting, the equilibrium configuration of the solid is the one given in \eqref{vev}. We can introduce an analogue of the mechanical deformations by taking
\be\label{pert}
\phi^I = \hat \phi^I + \pi^I\,.
\ee
Under this deformation the matrix $\mathcal I^{IJ}$ transforms as $\mathcal I^{IJ}\to\hat{\mathcal I}^{IJ}+\partial^I\pi^J+\partial^J\pi^I$ with the equilibrium value $\hat{\mathcal I}^{IJ}=\delta^{IJ}$. This provides a natural identification of the displacement tensor $u_{ij}$, given in \eqref{strain_tensor}, in terms of the symmetric combination of the derivatives of the scalar field perturbations $\pi^I$ as
\be
u^{IJ} \equiv \frac{1}{2}\left({\cal I}^{IJ} - \hat {\cal I}^{IJ}\right)\,.
\ee
It is clear now that this deformation will affect the energy density and the full stress tensor of the system. The notion of the shear modulus is therefore easy to identify in the EFT: a small transverse traceless shear deformation $u^{IJ}_{\text{T}}$ induces a small non-zero transverse traceless component of the $\phi^I$ stress tensor. This is the internal stress generated by such a strain, and is the one that has to be applied by an external action in order to sustain the given shear deformation constant in time. One can then obtain easily the elastic response by calculating 
\be
\delta T^{ij}_{\text{T}} =  T^{ij}_{\text{T}} - \hat T^{ij}_{\text{T}}
\ee
from the full stress-energy tensor of the scalar fields. The resulting expression can be put in the form\footnote{At the level of perturbations one can ignore the difference between $I$ and $i$ indices.}
\be
\delta T^{ij}_{\text{T}}=\mu\left({\cal I}^{ij} - \hat {\cal I}^{ij}\right)_{\text{T}}=2\mu \,  u^{ij}_{\text{T}}\,
\ee
with 
\be
\mu= \frac23 \left(X \, F_X+4YF_Y\right) \,.
\ee
A non-trivial cross-check that this is an appropriate description of a solid is to reproduce the linear dispersion relations for the phonons, $\omega = c_{L,\,T} \, k$,
with the known relations between the elastic moduli and the propagation speed of the transverse and longitudinal sound waves \cite{landau7,Lubensky}:
\begin{equation}\label{sound_non_rel}
c_L^2\,=\,\frac{\kappa\,+\,\frac{4}{3}\,\mu}{\varepsilon}\,,\qquad c_T^2\,=\,\frac{\mu}{\varepsilon}~.
\end{equation}
In a relativistic theory, one can expect that the denominator in \eqref{sound_non_rel} is not just given by the energy density $\varepsilon$. We can easily derive the relativistic version of these expressions by perturbing the background configuration \eqref{vev} with a spacetime dependent small perturbations $\pi^I$ defined in \eqref{pert}. They trivially split into transverse and longitudinal phonon sectors, the transverse one obeying $\partial_i\pi^i=0$. A straightforward calculation gives for the transverse sound speed
\be
c_T^2 = \frac{X \, F_X + 4\, Y \, F_Y }{ X \, F_X + 2\, Y \, F_Y + 3\, Z \, F_Z }
= \,\frac{\mu}{ \varepsilon+p}\,.
\ee
The expression for the longitudinal sound speed reproduces $c_L^2=(\kappa+\frac{4}{3}\mu)/(\varepsilon +p)$. These relations match with the well-known results for solids given in \eqref{sound_rel} and thus represents a zeroth order cross-check that the actions of the type \eqref{lagr} do really describe solids. In addition, we see that in the denominators the factor $\varepsilon+p$ emerges, as expected for relativistic systems.

Let us also emphasize that the fact that we were able to reproduce the correct formulas for the speeds from the simple solid effective Lagrangian means that these relations formally hold only at zero temperature. In other words, they are valid for the single-particle phonon excitations around the QFT vacuum state of the $\phi^I$ scalar fields. This is, of course, expected because the linear elastic response is reversible and non-dissipative, and therefore it must be derivable from a zero-temperature Lagrangian.
Thermal and other dissipative effects that affect the phonon dispersion relations can be incorporated in the EFT description, but this is not the topic of the present paper. Let us keep in mind, though, that in the holographic realizations presented in the main text, the dissipative effects  are indeed present.

\section{Numerical details of QNM computation}\label{app:numerics}

In this section we present some details regarding the numerics of the quasi-normal mode computation presented in the main text. To determine the quasi-normal modes which correspond to poles of the corresponding retarded Green's function in the dual field theory, we have to solve \eqref{eomU} and \eqref{eomB} subject to in-falling boundary conditions at the horizon. Moreover, by definition, the non-normalizable modes of the fields $B_i$ and $U_i$ have to vanish. 

Due to numerical efficiency we use instead directly the fluctuations $h_{ij}$ and $\varphi^I$. We perform a Fourier transformation for the fluctuation fields with respect to time and the spatial field theory directions. Moreover, along the radial direction the fluctuation fields are represented by Chebyshev polynomials.

Without loss of generality, the $x$-component of the momentum $k$ can be set to zero due to the $SO(2)$ rotational symmetry of the background in the $(x,y)$-plane. The relevant transverse coupled fluctuations are $h_{xu}, h_{xy}, h_{tx}$ and $\varphi^1.$ In the following we choose a gauge such that $h_{xu}=0$, hence giving rise to a constraint equation.

The in-falling horizon conditions are automatically  implemented by using in-going Eddington-Finkelstein coordinates. Moreover, by redefining the fields $h_{xy}, h_{tx}$ and $\varphi^1$ by $h_{xy}(z) = z \, \tilde{h}_{xy}(z), h_{tx}(z) = z \, \tilde{h}_{tx}(z)$ and $\varphi^1(z) = z^3 \tilde{\varphi}^1(z),$ respectively, and solving for $\tilde{h}_{xy}(z), \tilde{h}_{tx}(z)$ and $\tilde{\varphi}^1(z)$, we ensure that the non-normalizable modes vanish since only regular solutions can be represented on a Chebyshev grid. 

The problem of finding the quasi-normal modes and their associated solutions can be reformulated in terms of a generalized eigenvalue / eigenvector problem. \footnote{Note that despite using Eddington-Finkelstein coordinates, there are still $\omega^2$ terms appearing in the equations of motion for the fluctuations. However, by introducing auxiliary fields we reduce the generalised eigenvalue problem to a linear one. For more details see \cite{Ammon:2017ded}.} In particular, we solve for the three second-order equation of motion of $\tilde{h}_{xy}, \tilde{h}_{tx}$ and $\tilde{\varphi}^1$ and check afterwards whether the constraint equation is satisfied.

In particular, all QNM data in this project are calculated with at least 50 or 70 grid points and 50 digit precision. We checked the convergence of the quasi-normal modes as well as how good the equations of motion and the constraint are satisfied. 

\begin{figure}[htp]
\begin{center}
\includegraphics[width=0.32\textwidth]{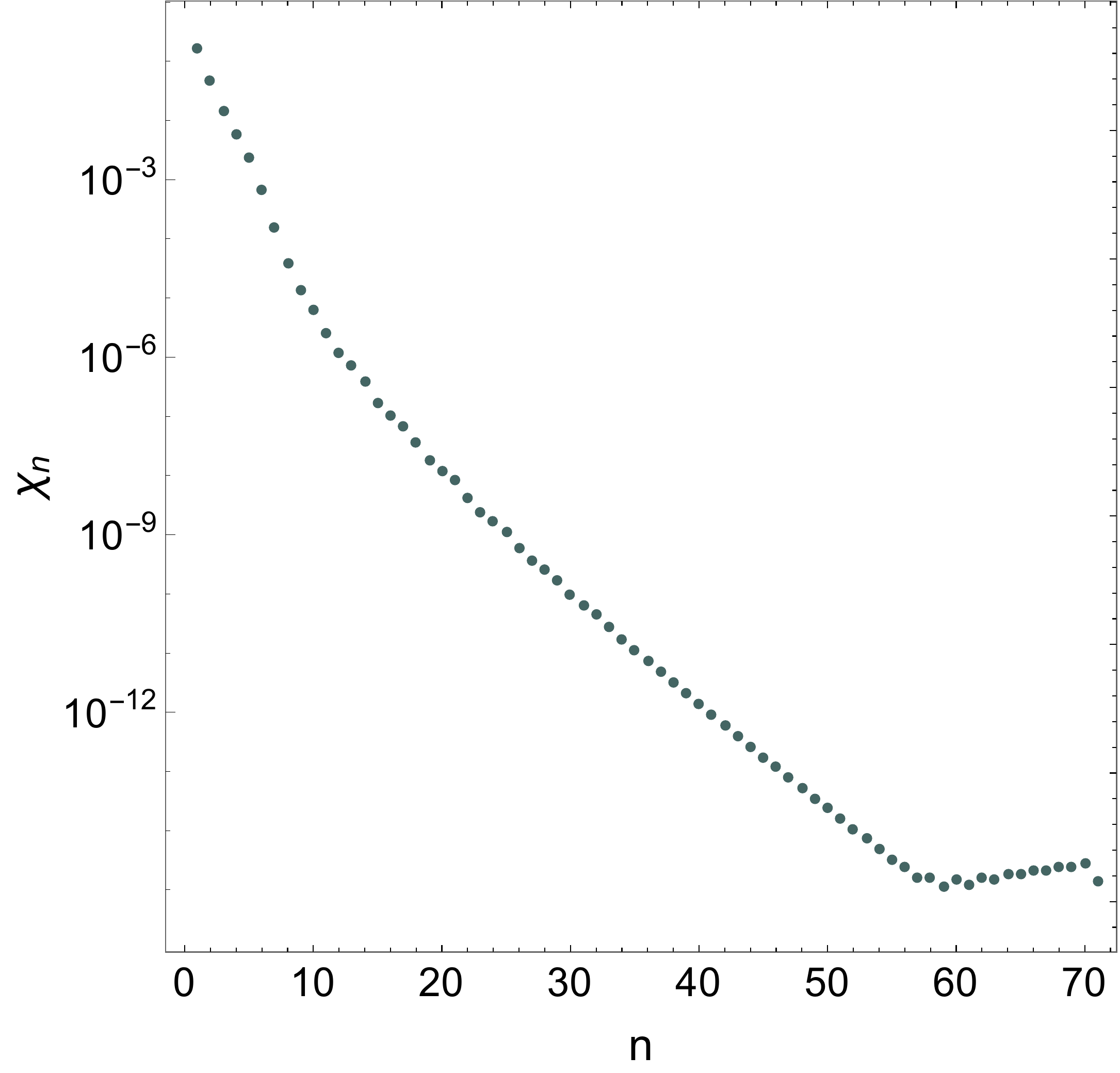}
\includegraphics[width=0.32\textwidth]{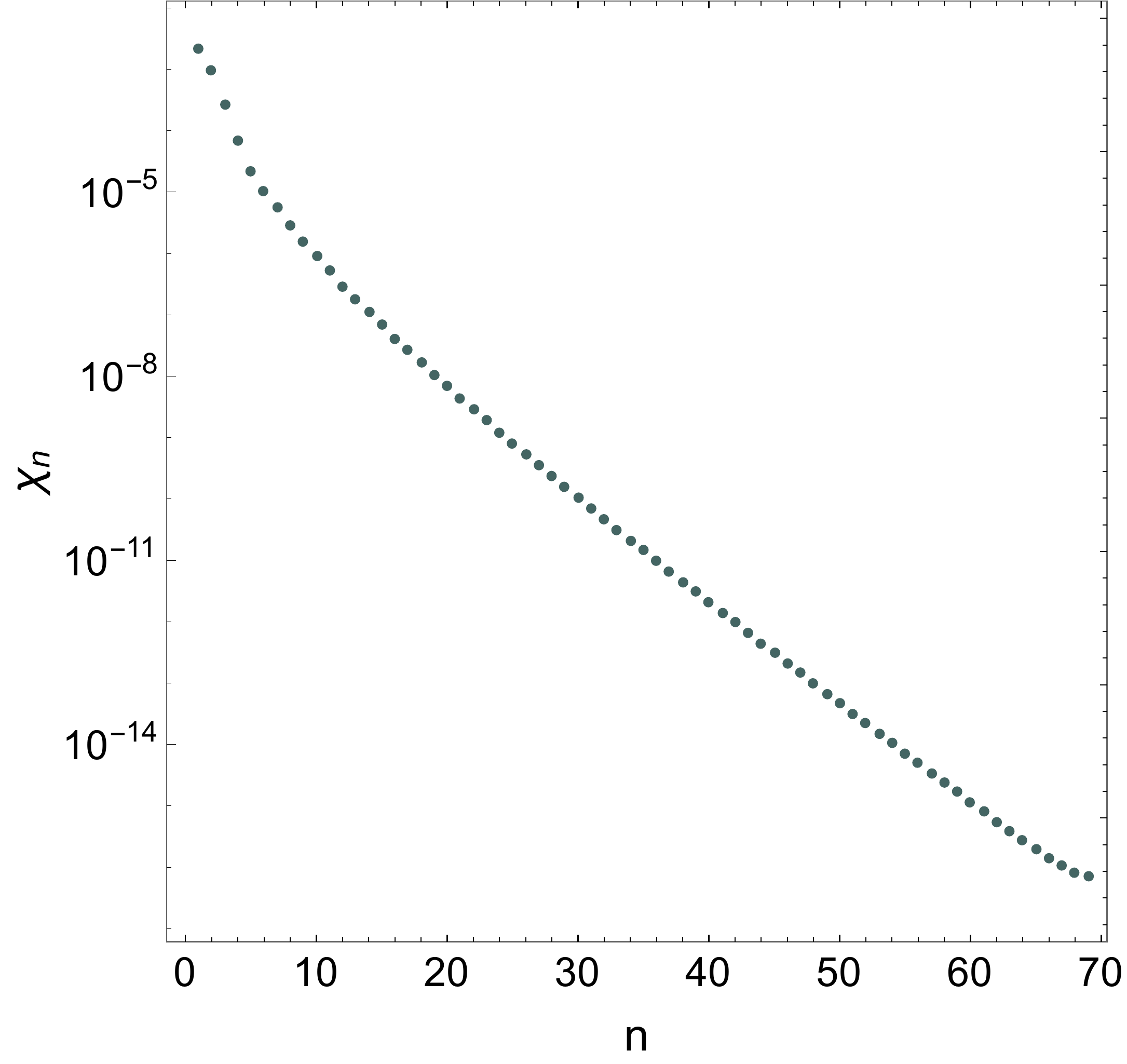}
\includegraphics[width=0.32\textwidth]{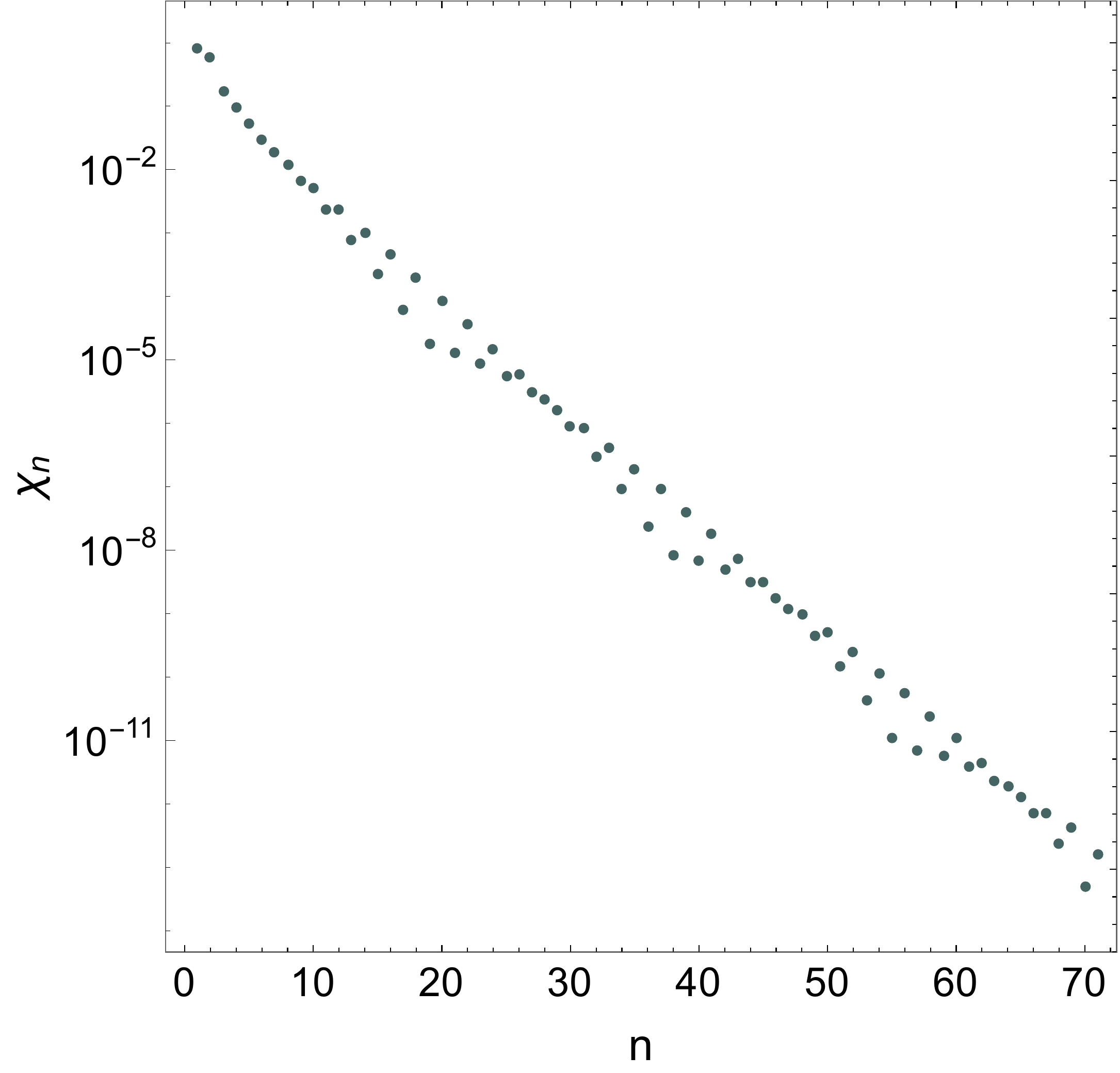}
  \caption{\label{fig:conv}Convergence plots of the functions $\tilde{h}_{xy}, \tilde{h}_{tx}$ and $\tilde{\varphi}^1$ for the lowest QNM for $k/T= 0.14,$ $\beta=60$ and $T/m=0.16$}
\end{center}
\end{figure}
In Fig. \ref{fig:conv} we present convergence plots for the lowest QNM for the following parameters: $\beta=60, T/m=0.16$ and $k/T= 0.14$. For this computation we used 70 grid points. Moreover, for this case we evaluated the equations of motion and the constraint on the solution using 100 regularly spaced points, finding the square norm to be $10^{-7}$ or less for each of the equations of motions as well as for the constraint.
\bibliographystyle{JHEP}
\bibliography{ArXivV4}

\providecommand{\href}[2]{#2}\begingroup\raggedright\begin{thebibliography}{10}

\bibitem{Leutwyler:1993gf}
H.~Leutwyler, \emph{{Nonrelativistic effective Lagrangians}},
  \href{http://dx.doi.org/10.1103/PhysRevD.49.3033}{\emph{Phys. Rev.} {\bf D49}
  (1994) 3033--3043}, [\href{https://arxiv.org/abs/hep-ph/9311264}{{\tt
  hep-ph/9311264}}].

\bibitem{Leutwyler:1996er}
H.~Leutwyler, \emph{{Phonons as goldstone bosons}}, {\emph{Helv. Phys. Acta}
  {\bf 70} (1997) 275--286}, [\href{https://arxiv.org/abs/hep-ph/9609466}{{\tt
  hep-ph/9609466}}].

\bibitem{Lubensky}
P.~M. Chaikin and T.~C. Lubensky, \emph{Principles of Condensed Matter
  Physics}.
\newblock Cambridge University Press, 1995,
  \href{http://dx.doi.org/10.1017/CBO9780511813467}{10.1017/CBO9780511813467}.

\bibitem{Gruner:1994zz}
G.~Gruner, \emph{{The dynamics of spin-density waves}},
  \href{http://dx.doi.org/10.1103/RevModPhys.66.1}{\emph{Rev. Mod. Phys.} {\bf
  66} (1994) 1--24}.

\bibitem{Maldacena:1997re}
J.~M. Maldacena, \emph{{The Large N limit of superconformal field theories and
  supergravity}}, \href{http://dx.doi.org/10.1023/A:1026654312961}{\emph{Int.
  J. Theor. Phys.} {\bf 38} (1999) 1113--1133},
  [\href{https://arxiv.org/abs/hep-th/9711200}{{\tt hep-th/9711200}}].

\bibitem{Vegh:2013sk}
D.~Vegh, \emph{{Holography without translational symmetry}},
  \href{https://arxiv.org/abs/1301.0537}{{\tt 1301.0537}}.

\bibitem{Blake:2013bqa}
M.~Blake and D.~Tong, \emph{{Universal Resistivity from Holographic Massive
  Gravity}}, \href{http://dx.doi.org/10.1103/PhysRevD.88.106004}{\emph{Phys.
  Rev.} {\bf D88} (2013) 106004}, [\href{https://arxiv.org/abs/1308.4970}{{\tt
  1308.4970}}].

\bibitem{Andrade:2013gsa}
T.~Andrade and B.~Withers, \emph{{A simple holographic model of momentum
  relaxation}}, \href{http://dx.doi.org/10.1007/JHEP05(2014)101}{\emph{JHEP}
  {\bf 05} (2014) 101}, [\href{https://arxiv.org/abs/1311.5157}{{\tt
  1311.5157}}].

\bibitem{Baggioli:2014roa}
M.~Baggioli and O.~Pujolas, \emph{{Electron-Phonon Interactions,
  Metal-Insulator Transitions, and Holographic Massive Gravity}},
  \href{http://dx.doi.org/10.1103/PhysRevLett.114.251602}{\emph{Phys. Rev.
  Lett.} {\bf 114} (2015) 251602}, [\href{https://arxiv.org/abs/1411.1003}{{\tt
  1411.1003}}].

\bibitem{Alberte:2015isw}
L.~Alberte, M.~Baggioli, A.~Khmelnitsky and O.~Pujolas, \emph{{Solid Holography
  and Massive Gravity}},
  \href{http://dx.doi.org/10.1007/JHEP02(2016)114}{\emph{JHEP} {\bf 02} (2016)
  114}, [\href{https://arxiv.org/abs/1510.09089}{{\tt 1510.09089}}].

\bibitem{Alberte:2016xja}
L.~Alberte, M.~Baggioli and O.~Pujolas, \emph{{Viscosity bound violation in
  holographic solids and the viscoelastic response}},
  \href{http://dx.doi.org/10.1007/JHEP07(2016)074}{\emph{JHEP} {\bf 07} (2016)
  074}, [\href{https://arxiv.org/abs/1601.03384}{{\tt 1601.03384}}].

\bibitem{landau7}
L.~D. Landau and E.~M. Lifshitz, \emph{Course of Theoretical Physics, Vol.
  7,Theory of Elasticity}.
\newblock Pergamon Press, 1970.

\bibitem{Delacretaz:2017zxd}
L.~V. Delacrétaz, B.~Goutéraux, S.~A. Hartnoll and A.~Karlsson,
  \emph{{Hydrodynamic transport in fluctuating charge density waves}},
  \href{https://arxiv.org/abs/1702.05104}{{\tt 1702.05104}}.

\bibitem{Dubovsky:2011sj}
S.~Dubovsky, L.~Hui, A.~Nicolis and D.~T. Son, \emph{{Effective field theory
  for hydrodynamics: thermodynamics, and the derivative expansion}},
  \href{http://dx.doi.org/10.1103/PhysRevD.85.085029}{\emph{Phys. Rev.} {\bf
  D85} (2012) 085029}, [\href{https://arxiv.org/abs/1107.0731}{{\tt
  1107.0731}}].

\bibitem{Endlich:2012pz}
S.~Endlich, A.~Nicolis and J.~Wang, \emph{{Solid Inflation}},
  \href{http://dx.doi.org/10.1088/1475-7516/2013/10/011}{\emph{JCAP} {\bf 1310}
  (2013) 011}, [\href{https://arxiv.org/abs/1210.0569}{{\tt 1210.0569}}].

\bibitem{Nicolis:2013lma}
A.~Nicolis, R.~Penco and R.~A. Rosen, \emph{{Relativistic Fluids, Superfluids,
  Solids and Supersolids from a Coset Construction}},
  \href{http://dx.doi.org/10.1103/PhysRevD.89.045002}{\emph{Phys. Rev.} {\bf
  D89} (2014) 045002}, [\href{https://arxiv.org/abs/1307.0517}{{\tt
  1307.0517}}].

\bibitem{Davison:2014lua}
R.~A. Davison and B.~Goutéraux, \emph{{Momentum dissipation and effective
  theories of coherent and incoherent transport}},
  \href{http://dx.doi.org/10.1007/JHEP01(2015)039}{\emph{JHEP} {\bf 01} (2015)
  039}, [\href{https://arxiv.org/abs/1411.1062}{{\tt 1411.1062}}].

\bibitem{Kim:2014bza}
K.-Y. Kim, K.~K. Kim, Y.~Seo and S.-J. Sin, \emph{{Coherent/incoherent metal
  transition in a holographic model}},
  \href{http://dx.doi.org/10.1007/JHEP12(2014)170}{\emph{JHEP} {\bf 12} (2014)
  170}, [\href{https://arxiv.org/abs/1409.8346}{{\tt 1409.8346}}].

\bibitem{Baggioli:2017ojd}
M.~Baggioli and W.-J. Li, \emph{{Diffusivities bounds and chaos in holographic
  Horndeski theories}},
  \href{http://dx.doi.org/10.1007/JHEP07(2017)055}{\emph{JHEP} {\bf 07} (2017)
  055}, [\href{https://arxiv.org/abs/1705.01766}{{\tt 1705.01766}}].

\bibitem{Hartnoll:2014lpa}
S.~A. Hartnoll, \emph{{Theory of universal incoherent metallic transport}},
  \href{http://dx.doi.org/10.1038/nphys3174}{\emph{Nature Phys.} {\bf 11}
  (2015) 54}, [\href{https://arxiv.org/abs/1405.3651}{{\tt 1405.3651}}].

\bibitem{Grozdanov:2015qia}
S.~Grozdanov, A.~Lucas, S.~Sachdev and K.~Schalm, \emph{{Absence of
  disorder-driven metal-insulator transitions in simple holographic models}},
  \href{http://dx.doi.org/10.1103/PhysRevLett.115.221601}{\emph{Phys. Rev.
  Lett.} {\bf 115} (2015) 221601},
  [\href{https://arxiv.org/abs/1507.00003}{{\tt 1507.00003}}].

\bibitem{Davison:2015taa}
R.~A. Davison, B.~Goutéraux and S.~A. Hartnoll, \emph{{Incoherent transport in
  clean quantum critical metals}},
  \href{http://dx.doi.org/10.1007/JHEP10(2015)112}{\emph{JHEP} {\bf 10} (2015)
  112}, [\href{https://arxiv.org/abs/1507.07137}{{\tt 1507.07137}}].

\bibitem{Grozdanov:2015djs}
S.~Grozdanov, A.~Lucas and K.~Schalm, \emph{{Incoherent thermal transport from
  dirty black holes}},
  \href{http://dx.doi.org/10.1103/PhysRevD.93.061901}{\emph{Phys. Rev.} {\bf
  D93} (2016) 061901}, [\href{https://arxiv.org/abs/1511.05970}{{\tt
  1511.05970}}].

\bibitem{Blake:2016sud}
M.~Blake, \emph{{Universal Diffusion in Incoherent Black Holes}},
  \href{http://dx.doi.org/10.1103/PhysRevD.94.086014}{\emph{Phys. Rev.} {\bf
  D94} (2016) 086014}, [\href{https://arxiv.org/abs/1604.01754}{{\tt
  1604.01754}}].

\bibitem{Baggioli:2016pia}
M.~Baggioli, B.~Goutéraux, E.~Kiritsis and W.-J. Li, \emph{{Higher derivative
  corrections to incoherent metallic transport in holography}},
  \href{http://dx.doi.org/10.1007/JHEP03(2017)170}{\emph{JHEP} {\bf 03} (2017)
  170}, [\href{https://arxiv.org/abs/1612.05500}{{\tt 1612.05500}}].

\bibitem{Kim:2017dgz}
K.-Y. Kim and C.~Niu, \emph{{Diffusion and Butterfly Velocity at Finite
  Density}}, \href{http://dx.doi.org/10.1007/JHEP06(2017)030}{\emph{JHEP} {\bf
  06} (2017) 030}, [\href{https://arxiv.org/abs/1704.00947}{{\tt 1704.00947}}].

\bibitem{Blake:2017qgd}
M.~Blake, R.~A. Davison and S.~Sachdev, \emph{{Thermal diffusivity and chaos in
  metals without quasiparticles}},
  \href{https://arxiv.org/abs/1705.07896}{{\tt 1705.07896}}.

\bibitem{Lucas:2015vna}
A.~Lucas, \emph{{Conductivity of a strange metal: from holography to memory
  functions}}, \href{http://dx.doi.org/10.1007/JHEP03(2015)071}{\emph{JHEP}
  {\bf 03} (2015) 071}, [\href{https://arxiv.org/abs/1501.05656}{{\tt
  1501.05656}}].

\bibitem{Davison:2015bea}
R.~A. Davison and B.~Goutéraux, \emph{{Dissecting holographic
  conductivities}},
  \href{http://dx.doi.org/10.1007/JHEP09(2015)090}{\emph{JHEP} {\bf 09} (2015)
  090}, [\href{https://arxiv.org/abs/1505.05092}{{\tt 1505.05092}}].

\bibitem{Lucas:2015lna}
A.~Lucas, \emph{{Hydrodynamic transport in strongly coupled disordered quantum
  field theories}},
  \href{http://dx.doi.org/10.1088/1367-2630/17/11/113007}{\emph{New J. Phys.}
  {\bf 17} (2015) 113007}, [\href{https://arxiv.org/abs/1506.02662}{{\tt
  1506.02662}}].

\bibitem{Baggioli:2016oqk}
M.~Baggioli and O.~Pujolas, \emph{{On holographic disorder-driven
  metal-insulator transitions}},
  \href{http://dx.doi.org/10.1007/JHEP01(2017)040}{\emph{JHEP} {\bf 01} (2017)
  040}, [\href{https://arxiv.org/abs/1601.07897}{{\tt 1601.07897}}].

\bibitem{Davison:2013jba}
R.~A. Davison, \emph{{Momentum relaxation in holographic massive gravity}},
  \href{http://dx.doi.org/10.1103/PhysRevD.88.086003}{\emph{Phys. Rev.} {\bf
  D88} (2013) 086003}, [\href{https://arxiv.org/abs/1306.5792}{{\tt
  1306.5792}}].

\bibitem{Blake:2013owa}
M.~Blake, D.~Tong and D.~Vegh, \emph{{Holographic Lattices Give the Graviton an
  Effective Mass}},
  \href{http://dx.doi.org/10.1103/PhysRevLett.112.071602}{\emph{Phys. Rev.
  Lett.} {\bf 112} (2014) 071602}, [\href{https://arxiv.org/abs/1310.3832}{{\tt
  1310.3832}}].

\bibitem{Davison:2013txa}
R.~A. Davison, K.~Schalm and J.~Zaanen, \emph{{Holographic duality and the
  resistivity of strange metals}},
  \href{http://dx.doi.org/10.1103/PhysRevB.89.245116}{\emph{Phys. Rev.} {\bf
  B89} (2014) 245116}, [\href{https://arxiv.org/abs/1311.2451}{{\tt
  1311.2451}}].

\bibitem{Donos:2013eha}
A.~Donos and J.~P. Gauntlett, \emph{{Holographic Q-lattices}},
  \href{http://dx.doi.org/10.1007/JHEP04(2014)040}{\emph{JHEP} {\bf 04} (2014)
  040}, [\href{https://arxiv.org/abs/1311.3292}{{\tt 1311.3292}}].

\bibitem{Gouteraux:2014hca}
B.~Goutéraux, \emph{{Charge transport in holography with momentum
  dissipation}}, \href{http://dx.doi.org/10.1007/JHEP04(2014)181}{\emph{JHEP}
  {\bf 04} (2014) 181}, [\href{https://arxiv.org/abs/1401.5436}{{\tt
  1401.5436}}].

\bibitem{Donos:2014uba}
A.~Donos and J.~P. Gauntlett, \emph{{Novel metals and insulators from
  holography}}, \href{http://dx.doi.org/10.1007/JHEP06(2014)007}{\emph{JHEP}
  {\bf 06} (2014) 007}, [\href{https://arxiv.org/abs/1401.5077}{{\tt
  1401.5077}}].

\bibitem{Amoretti:2014zha}
A.~Amoretti, A.~Braggio, N.~Maggiore, N.~Magnoli and D.~Musso,
  \emph{{Thermo-electric transport in gauge/gravity models with momentum
  dissipation}}, \href{http://dx.doi.org/10.1007/JHEP09(2014)160}{\emph{JHEP}
  {\bf 09} (2014) 160}, [\href{https://arxiv.org/abs/1406.4134}{{\tt
  1406.4134}}].

\bibitem{Donos:2014oha}
A.~Donos, B.~Goutéraux and E.~Kiritsis, \emph{{Holographic Metals and
  Insulators with Helical Symmetry}},
  \href{http://dx.doi.org/10.1007/JHEP09(2014)038}{\emph{JHEP} {\bf 09} (2014)
  038}, [\href{https://arxiv.org/abs/1406.6351}{{\tt 1406.6351}}].

\bibitem{Horowitz:2012ky}
G.~T. Horowitz, J.~E. Santos and D.~Tong, \emph{{Optical Conductivity with
  Holographic Lattices}},
  \href{http://dx.doi.org/10.1007/JHEP07(2012)168}{\emph{JHEP} {\bf 07} (2012)
  168}, [\href{https://arxiv.org/abs/1204.0519}{{\tt 1204.0519}}].

\bibitem{Baggioli:2016rdj}
M.~Baggioli, \emph{{Gravity, holography and applications to condensed matter}}.
\newblock PhD thesis, Barcelona U., 2016.
\newblock \href{https://arxiv.org/abs/1610.02681}{{\tt 1610.02681}}.

\bibitem{Baggioli:2016oju}
M.~Baggioli and O.~Pujolas, \emph{{On Effective Holographic Mott Insulators}},
  \href{http://dx.doi.org/10.1007/JHEP12(2016)107}{\emph{JHEP} {\bf 12} (2016)
  107}, [\href{https://arxiv.org/abs/1604.08915}{{\tt 1604.08915}}].

\bibitem{Amoretti:2016cad}
A.~Amoretti, M.~Baggioli, N.~Magnoli and D.~Musso, \emph{{Chasing the cuprates
  with dilatonic dyons}},
  \href{http://dx.doi.org/10.1007/JHEP06(2016)113}{\emph{JHEP} {\bf 06} (2016)
  113}, [\href{https://arxiv.org/abs/1603.03029}{{\tt 1603.03029}}].

\bibitem{Amado:2013xya}
I.~Amado, D.~Arean, A.~Jimenez-Alba, K.~Landsteiner, L.~Melgar and I.~S.
  Landea, \emph{{Holographic Type II Goldstone bosons}},
  \href{http://dx.doi.org/10.1007/JHEP07(2013)108}{\emph{JHEP} {\bf 07} (2013)
  108}, [\href{https://arxiv.org/abs/1302.5641}{{\tt 1302.5641}}].

\bibitem{Argurio:2015via}
R.~Argurio, A.~Marzolla, A.~Mezzalira and D.~Naegels, \emph{{Note on
  holographic nonrelativistic Goldstone bosons}},
  \href{http://dx.doi.org/10.1103/PhysRevD.92.066009}{\emph{Phys. Rev.} {\bf
  D92} (2015) 066009}, [\href{https://arxiv.org/abs/1507.00211}{{\tt
  1507.00211}}].

\bibitem{Esposito:2016ria}
A.~Esposito, S.~Garcia-Saenz and R.~Penco, \emph{{First sound in holographic
  superfluids at zero temperature}},
  \href{http://dx.doi.org/10.1007/JHEP12(2016)136}{\emph{JHEP} {\bf 12} (2016)
  136}, [\href{https://arxiv.org/abs/1606.03104}{{\tt 1606.03104}}].

\bibitem{Argurio:2015wgr}
R.~Argurio, A.~Marzolla, A.~Mezzalira and D.~Musso, \emph{{Analytic
  pseudo-Goldstone bosons}},
  \href{http://dx.doi.org/10.1007/JHEP03(2016)012}{\emph{JHEP} {\bf 03} (2016)
  012}, [\href{https://arxiv.org/abs/1512.03750}{{\tt 1512.03750}}].

\bibitem{Argurio:2016xih}
R.~Argurio, G.~Giribet, A.~Marzolla, D.~Naegels and J.~A. Sierra-Garcia,
  \emph{{Holographic Ward identities for symmetry breaking in two dimensions}},
  \href{http://dx.doi.org/10.1007/JHEP04(2017)007}{\emph{JHEP} {\bf 04} (2017)
  007}, [\href{https://arxiv.org/abs/1612.00771}{{\tt 1612.00771}}].

\bibitem{Amoretti:2016bxs}
A.~Amoretti, D.~Areán, R.~Argurio, D.~Musso and L.~A. Pando~Zayas, \emph{{A
  holographic perspective on phonons and pseudo-phonons}},
  \href{http://dx.doi.org/10.1007/JHEP05(2017)051}{\emph{JHEP} {\bf 05} (2017)
  051}, [\href{https://arxiv.org/abs/1611.09344}{{\tt 1611.09344}}].

\bibitem{Nakamura:2009tf}
S.~Nakamura, H.~Ooguri and C.-S. Park, \emph{{Gravity Dual of Spatially
  Modulated Phase}},
  \href{http://dx.doi.org/10.1103/PhysRevD.81.044018}{\emph{Phys. Rev.} {\bf
  D81} (2010) 044018}, [\href{https://arxiv.org/abs/0911.0679}{{\tt
  0911.0679}}].

\bibitem{Ooguri:2010kt}
H.~Ooguri and C.-S. Park, \emph{{Holographic End-Point of Spatially Modulated
  Phase Transition}},
  \href{http://dx.doi.org/10.1103/PhysRevD.82.126001}{\emph{Phys. Rev.} {\bf
  D82} (2010) 126001}, [\href{https://arxiv.org/abs/1007.3737}{{\tt
  1007.3737}}].

\bibitem{Aperis:2010cd}
A.~Aperis, P.~Kotetes, E.~Papantonopoulos, G.~Siopsis, P.~Skamagoulis and
  G.~Varelogiannis, \emph{{Holographic Charge Density Waves}},
  \href{http://dx.doi.org/10.1016/j.physletb.2011.06.092}{\emph{Phys. Lett.}
  {\bf B702} (2011) 181--185}, [\href{https://arxiv.org/abs/1009.6179}{{\tt
  1009.6179}}].

\bibitem{Ooguri:2010xs}
H.~Ooguri and C.-S. Park, \emph{{Spatially Modulated Phase in Holographic
  Quark-Gluon Plasma}},
  \href{http://dx.doi.org/10.1103/PhysRevLett.106.061601}{\emph{Phys. Rev.
  Lett.} {\bf 106} (2011) 061601}, [\href{https://arxiv.org/abs/1011.4144}{{\tt
  1011.4144}}].

\bibitem{Donos:2011bh}
A.~Donos and J.~P. Gauntlett, \emph{{Holographic striped phases}},
  \href{http://dx.doi.org/10.1007/JHEP08(2011)140}{\emph{JHEP} {\bf 08} (2011)
  140}, [\href{https://arxiv.org/abs/1106.2004}{{\tt 1106.2004}}].

\bibitem{Bergman:2011rf}
O.~Bergman, N.~Jokela, G.~Lifschytz and M.~Lippert, \emph{{Striped instability
  of a holographic Fermi-like liquid}},
  \href{http://dx.doi.org/10.1007/JHEP10(2011)034}{\emph{JHEP} {\bf 10} (2011)
  034}, [\href{https://arxiv.org/abs/1106.3883}{{\tt 1106.3883}}].

\bibitem{Donos:2011qt}
A.~Donos, J.~P. Gauntlett and C.~Pantelidou, \emph{{Spatially modulated
  instabilities of magnetic black branes}},
  \href{http://dx.doi.org/10.1007/JHEP01(2012)061}{\emph{JHEP} {\bf 01} (2012)
  061}, [\href{https://arxiv.org/abs/1109.0471}{{\tt 1109.0471}}].

\bibitem{Donos:2012wi}
A.~Donos and J.~P. Gauntlett, \emph{{Black holes dual to helical current
  phases}}, \href{http://dx.doi.org/10.1103/PhysRevD.86.064010}{\emph{Phys.
  Rev.} {\bf D86} (2012) 064010}, [\href{https://arxiv.org/abs/1204.1734}{{\tt
  1204.1734}}].

\bibitem{Ammon:2016szz}
M.~Ammon, J.~Leiber and R.~P. Macedo, \emph{{Phase diagram of 4D field theories
  with chiral anomaly from holography}},
  \href{http://dx.doi.org/10.1007/JHEP03(2016)164}{\emph{JHEP} {\bf 03} (2016)
  164}, [\href{https://arxiv.org/abs/1601.02125}{{\tt 1601.02125}}].

\bibitem{Delacretaz:2016ivq}
L.~V. Delacrétaz, B.~Goutéraux, S.~A. Hartnoll and A.~Karlsson, \emph{{Bad
  Metals from Fluctuating Density Waves}},
  \href{https://arxiv.org/abs/1612.04381}{{\tt 1612.04381}}.

\bibitem{Nicolis:2015sra}
A.~Nicolis, R.~Penco, F.~Piazza and R.~Rattazzi, \emph{{Zoology of condensed
  matter: Framids, ordinary stuff, extra-ordinary stuff}},
  \href{http://dx.doi.org/10.1007/JHEP06(2015)155}{\emph{JHEP} {\bf 06} (2015)
  155}, [\href{https://arxiv.org/abs/1501.03845}{{\tt 1501.03845}}].

\bibitem{Jokela:2017ltu}
N.~Jokela, M.~Jarvinen and M.~Lippert, \emph{{Holographic pinning}},
  \href{https://arxiv.org/abs/1708.07837}{{\tt 1708.07837}}.

\bibitem{Andrade:2017cnc}
T.~Andrade, M.~Baggioli, A.~Krikun and N.~Poovuttikul, \emph{{Pinning of
  longitudinal phonons in holographic helical crystals}},
  \href{https://arxiv.org/abs/1708.08306}{{\tt 1708.08306}}.

\bibitem{KADANOFF}
L.~P. Kadanoff and P.~C. Martin, \emph{Hydrodynamic equations and correlation
  functions},
  \href{http://dx.doi.org/http://dx.doi.org/10.1016/0003-4916(63)90078-2}{\emph{Annals
  of Physics} {\bf 24} (1963) 419 -- 469}.

\bibitem{deHaro:2000vlm}
S.~de~Haro, S.~N. Solodukhin and K.~Skenderis, \emph{{Holographic
  reconstruction of space-time and renormalization in the AdS / CFT
  correspondence}},
  \href{http://dx.doi.org/10.1007/s002200100381}{\emph{Commun. Math. Phys.}
  {\bf 217} (2001) 595--622}, [\href{https://arxiv.org/abs/hep-th/0002230}{{\tt
  hep-th/0002230}}].

\bibitem{Hartnoll:2009sz}
S.~A. Hartnoll, \emph{{Lectures on holographic methods for condensed matter
  physics}},
  \href{http://dx.doi.org/10.1088/0264-9381/26/22/224002}{\emph{Class. Quant.
  Grav.} {\bf 26} (2009) 224002}, [\href{https://arxiv.org/abs/0903.3246}{{\tt
  0903.3246}}].

\bibitem{Hartnoll:2016tri}
S.~A. Hartnoll, D.~M. Ramirez and J.~E. Santos, \emph{{Entropy production,
  viscosity bounds and bumpy black holes}},
  \href{http://dx.doi.org/10.1007/JHEP03(2016)170}{\emph{JHEP} {\bf 03} (2016)
  170}, [\href{https://arxiv.org/abs/1601.02757}{{\tt 1601.02757}}].

\bibitem{Burikham:2016roo}
P.~Burikham and N.~Poovuttikul, \emph{{Shear viscosity in holography and
  effective theory of transport without translational symmetry}},
  \href{http://dx.doi.org/10.1103/PhysRevD.94.106001}{\emph{Phys. Rev.} {\bf
  D94} (2016) 106001}, [\href{https://arxiv.org/abs/1601.04624}{{\tt
  1601.04624}}].

\bibitem{Policastro:2001yc}
G.~Policastro, D.~T. Son and A.~O. Starinets, \emph{{The Shear viscosity of
  strongly coupled N=4 supersymmetric Yang-Mills plasma}},
  \href{http://dx.doi.org/10.1103/PhysRevLett.87.081601}{\emph{Phys. Rev.
  Lett.} {\bf 87} (2001) 081601},
  [\href{https://arxiv.org/abs/hep-th/0104066}{{\tt hep-th/0104066}}].

\bibitem{Blake:2015epa}
M.~Blake, \emph{{Momentum relaxation from the fluid/gravity correspondence}},
  \href{http://dx.doi.org/10.1007/JHEP09(2015)010}{\emph{JHEP} {\bf 09} (2015)
  010}, [\href{https://arxiv.org/abs/1505.06992}{{\tt 1505.06992}}].

\bibitem{Jimenez-Alba:2014iia}
A.~Jimenez-Alba, K.~Landsteiner and L.~Melgar, \emph{{Anomalous magnetoresponse
  and the Stückelberg axion in holography}},
  \href{http://dx.doi.org/10.1103/PhysRevD.90.126004}{\emph{Phys. Rev.} {\bf
  D90} (2014) 126004}, [\href{https://arxiv.org/abs/1407.8162}{{\tt
  1407.8162}}].

\bibitem{Grozdanov:2016vgg}
S.~Grozdanov, N.~Kaplis and A.~O. Starinets, \emph{{From strong to weak
  coupling in holographic models of thermalization}},
  \href{http://dx.doi.org/10.1007/JHEP07(2016)151}{\emph{JHEP} {\bf 07} (2016)
  151}, [\href{https://arxiv.org/abs/1605.02173}{{\tt 1605.02173}}].

\bibitem{Hofman:2017vwr}
D.~M. Hofman and N.~Iqbal, \emph{{Generalized global symmetries and
  holography}},  \href{https://arxiv.org/abs/1707.08577}{{\tt 1707.08577}}.

\bibitem{Stephanov:2014dma}
M.~Stephanov, H.-U. Yee and Y.~Yin, \emph{{Collective modes of chiral kinetic
  theory in a magnetic field}},
  \href{http://dx.doi.org/10.1103/PhysRevD.91.125014}{\emph{Phys. Rev.} {\bf
  D91} (2015) 125014}, [\href{https://arxiv.org/abs/1501.00222}{{\tt
  1501.00222}}].

\bibitem{Gursoy:2013zxa}
U.~Gürsoy, S.~Lin and E.~Shuryak, \emph{{Instabilities near the QCD phase
  transition in the holographic models}},
  \href{http://dx.doi.org/10.1103/PhysRevD.88.105021}{\emph{Phys. Rev.} {\bf
  D88} (2013) 105021}, [\href{https://arxiv.org/abs/1309.0789}{{\tt
  1309.0789}}].

\bibitem{Janik:2016btb}
R.~A. Janik, J.~Jankowski and H.~Soltanpanahi, \emph{{Quasinormal modes and the
  phase structure of strongly coupled matter}},
  \href{http://dx.doi.org/10.1007/JHEP06(2016)047}{\emph{JHEP} {\bf 06} (2016)
  047}, [\href{https://arxiv.org/abs/1603.05950}{{\tt 1603.05950}}].

\bibitem{Esposito:2017qpj}
A.~Esposito, S.~Garcia-Saenz, A.~Nicolis and R.~Penco, \emph{{Conformal solids
  and holography}},  \href{https://arxiv.org/abs/1708.09391}{{\tt 1708.09391}}.

\bibitem{Kachru:2009xf}
S.~Kachru, A.~Karch and S.~Yaida, \emph{{Holographic Lattices, Dimers, and
  Glasses}}, \href{http://dx.doi.org/10.1103/PhysRevD.81.026007}{\emph{Phys.
  Rev.} {\bf D81} (2010) 026007}, [\href{https://arxiv.org/abs/0909.2639}{{\tt
  0909.2639}}].

\bibitem{Anninos:2013mfa}
D.~Anninos, T.~Anous, F.~Denef and L.~Peeters, \emph{{Holographic
  Vitrification}}, \href{http://dx.doi.org/10.1007/JHEP04(2015)027}{\emph{JHEP}
  {\bf 04} (2015) 027}, [\href{https://arxiv.org/abs/1309.0146}{{\tt
  1309.0146}}].

\bibitem{Baggioli:2015gsa}
M.~Baggioli and D.~K. Brattan, \emph{{Drag phenomena from holographic massive
  gravity}},
  \href{http://dx.doi.org/10.1088/1361-6382/34/1/015008}{\emph{Class. Quant.
  Grav.} {\bf 34} (2017) 015008}, [\href{https://arxiv.org/abs/1504.07635}{{\tt
  1504.07635}}].

\bibitem{Alberte:2017oqx}
L.~Alberte, M.~Ammon, M.~Baggioli, A.~Jiménez-Alba and O.~Pujolàs,
  \emph{{Holographic Phonons}},  \href{https://arxiv.org/abs/1711.03100}{{\tt
  1711.03100}}.

\bibitem{Ammon:2017ded}
M.~Ammon, M.~Kaminski, R.~Koirala, J.~Leiber and J.~Wu, \emph{{Quasinormal
  modes of charged magnetic black branes \& chiral magnetic transport}},
  \href{http://dx.doi.org/10.1007/JHEP04(2017)067}{\emph{JHEP} {\bf 04} (2017)
  067}, [\href{https://arxiv.org/abs/1701.05565}{{\tt 1701.05565}}].

\end{thebibliography}\endgroup
\end{document}